\documentclass[aps,pre,twocolumn,groupedaddress,superscriptaddress,amsmath,amssymb,amsfonts,floats,floatfix,balancelastpage,showpacs]{revtex4-2}

\usepackage{epsfig} 
\usepackage{psfrag}
\usepackage{braket}
\usepackage{color}
\usepackage[usenames,dvipsnames]{xcolor}
\usepackage{textcomp}
\usepackage{graphicx}
\usepackage{subfigure}
\usepackage{multirow}
\usepackage{latexsym}
\usepackage{url}

\usepackage{amssymb}
\usepackage{epstopdf}%This line makes .jpg figures into .pdf - please comment out if not required.

\definecolor{cream}{RGB}{222,217,201}

\begin{document}

\title{Building blocks of non-Euclidean ribbons: Size-controlled self-assembly via discrete frustrated particles}

\author{Douglas M. Hall}
\affiliation{Department of Polymer Science and Engineering, University of Massachusetts, Amherst, MA 01003, US}
\author{Mark J. Stevens}
\affiliation{Center for Integrated Nanotechnologies, Sandia National Laboratories, Albuquerque, NM 87185, USA.}
\author{Gregory M. Grason}
\affiliation{Department of Polymer Science and Engineering, University of Massachusetts, Amherst, MA 01003, US}

\begin{abstract}

Geometric frustration offers a pathway to soft matter self-assembly with controllable finite sizes.  While the understanding of frustration in soft matter assembly derives almost exclusively from continuum elastic descriptions, a current challenge is to understand the connection between microscopic physical properties of misfitting ``building blocks" and emergent assembly behavior at mesoscale.  We present and analyze a particle-based description of what is arguably the best studied example for frustrated soft matter assembly, negative-curvature ribbon assembly, observed in both assemblies of chiral surfactants and shape-frustrated nanoparticles.  Based on our particle model, known as  {\it saddle wedge monomers}, we numerically test the connection between microscopic shape and interactions of the misfitting subunits and the emergent behavior at the supra-particle scale, specifically focussing on the propagation and relaxation of inter-particle strains, the emergent role of extrinsic shape on frustrated ribbons and the equilibrium regime of finite width selection. Beyond the intuitive role of shape misfit, we show that self-limitation is critically dependent on the finite range of cohesive interactions, with larger size finite assemblies requiring increasing short-range interparticle forces. Additionally, we demonstrate that non-linearities arising from discrete particle interactions alter self-limiting behavior due to both strain-softening in shape-flattened assembly and partial yielding of highly strained bonds, which in turn may give rise to states of hierarchical, multidomain assembly.  Tracing the regimes of frustration-limited assembly to the specific microscopic features of misfitting particle shapes and interactions provides necessary guidance for translating the theory of size-programmable assembly into design of intentionally-frustrated colloidal particles.

\end{abstract}

\maketitle

\section{Introduction}

Geometric frustration (GF) occurs when the locally preferred ordering is incompatible with geometric constraints of extending that order throughout the assembly. \cite{Kleman1989, Sadoc2006} Canonically, GF is associate in bulk systems, where it requires the formation of extensive arrays of topological defects, as in polytetrahedral sphere packings~\cite{Nelson1989} or liquid crystal blue phases .\cite{Wright1989}  When a self-assembling system has GF, the presence of free boundaries on potentially forming finite-sized structures leads to distinct consequences and a range of exotic, scale-dependent thermodynamic behavior. \cite{grason2016perspective}  Notably, finite and sufficiently soft assemblies need not form defects as response to GF, which may instead manifest in a superextensive accumulation of intra-assembly stress that, in competition with the cohesive drive for assembly growth, may shape the assembly's equilbrium boundary and interior at length scales much larger than the subunit (e.g. macromolular or colloidal) dimensions. \cite{grason2016perspective, meiri2021cumulative}  Arguably, the most notable emergent behavior is the ability of the GF to determine the mesoscopic finite equilibrium size of assemblies. \cite{hagan2021}  
This basic paradigm has been explored in the context of a range of soft matter systems, from spherical assemblies of colloids~\cite{Schneider2005, meng2014elastic} and protein shells,~\cite{Mendoza2020} to twisted bundles of filamentous proteins or chiral fibers~\cite{Hall2016, Grason2020, Efrati2020} and chiral ribbons.~\cite{Aggeli2001, Achard2005, Ghafouri2005, Armon2014, Zhang2019, serafin2021frustrated}  The specific dependence on long-range gradients in intra-assembly stress and the resulting ability of thermodynamics to sense the mesoscopic size of assemblies distinguishes geometrically frustrated assembly (GFA) from other more familiar examples of size-selective assemblies, like amphiphillic micelles or self-closing, curvature limited shells and tubules.

Models for  size control in GFAs are generically predicated on continuum elastic descriptions of the super-extensive growth in assembly energy.~\cite{hagan2021}  These models argue that elastic energy accumulates with size up to an upper size limit, beyond which the assembly distorts away from the locally-preferred packing (at finite energy cost) to maintain extensive energetic growth with size.~\cite{meiri2021cumulative}  At these large sizes, frustration is not able to restrain the cohesive drive to larger size, and equilibrium assembly proceeds to unlimited size, known as {\it frustration escape}.  There are multiple possible structural modes of assembly:  elastic ``shape flattening'' of the preferred frustrated packing into an unfrustrated one;~\cite{Grason2020, Spivack_2022, tyukodi2021thermodynamic} ``filamentation'' into structures that remain finite in only a single direction of assembly but unlimited in others;~\cite{Schneider2005, Hall2016, Lenz2017} and incorporation of topological defect arrays that screen the far-field stresses responsible for cumulative frustration costs. \cite{Bruss2013, Hall2017, Paquay2017, Li2019}  
% \MJS{Is this screening or stress relief or both?}

At a conceptual level, the possibility of thermodynamic self-limitation as well as the existence of distinct modes of frustration escape that delimit the range of self-limitation for any given GFA is well established. The potential advantages posed by self-assembling systems that can ``sense'' their size at ranges that exceed the subunits themselves raises the possibility of intentionally engineering frustration into synthetically fabricated assemblies as a means to ``program" their assembly behavior.~\cite{Grason2017, Berengut2020, Tanjeem2022} In principle, recent progress in the synthesis of colloidal-scale particles with programmed shape can allow for tunable shape frustration that can more fully test the continuum theory description of size control. \cite{Glotzer2007, hueckel2021total} Notably, advances in DNA nanotechnology \cite{sigl2021programmable, hayakawa2022geometrically} as well as synthetic protein engineering \cite{hsia2016, Bale2016, King2014, wicky2022} allow for both careful design and control of  the shape frustration of self-assembling nanoscale units as well as new opportunities for programming the interactions to separately tune the strength of cohesion and costs associated with distinct modes of assembly deformation.  However, due to the primary reliance on continuum descriptions of GFA, several basic challenges remain to relate emergent thermodynamic behaviors in a particular system of self-assembling frustrated subunits.  In general, it remains to be understood which specific structural mechanisms are responsible for frustration escape in any particular system, and moreover, what are the size ranges, relative to the subunit dimensions, at which frustration may limit the thermodynamic assembly size.  Finally, beyond the continuum descriptions whose predictions rely on phenomenological constants of unknown value, how does the structure and thermodynamic range of self-limiting GFA depend on physical properties of the subunits themselves, their ill-fitting shapes, interactions and deformability?

%\MJS{there's an issue that theory is a 'membrane' theory, but the simulations use a 'nonmembrane' model. Not sure how to finesse this, but the first line caught me offguard so to speak.}
In this study, we focus on a particular well-studied model of GFA: crystalline membrane assemblies frustrated by preferred negative Gaussian curvature shapes.  Initial models of this type were motivated by observation of ribbon, or tape-like, assemblies of chiral amphiphile exhibiting  twisted, helicoidal ribbon morphologies with well-defined ribbon width. \cite{oda1999tuning, selinger2004shape, ziserman2011curvature}  In these assemblies preference for negative Gaussian curvature derives from the chirality of the molecules.~\cite{helfrich1988intrinsic} Helicoidal ribbon morphologies have also been observed in tetrahedral nanoparticles assembly. \cite{yan2019self, serafin2021frustrated}

The scale-dependent morphology of these structures has been described by a continuum elastic theory that accounts for growth of intra-ribbon strains of crystalline order in negatively curved ribbons, as well as the elastic (bending) preference for negative curvature.  The first model of this type was developed by Ghafouri and Bruinsma for chiral membranes, but has been subsequently elaborated on by several other studies. \cite{Ghafouri2005, Armon2014, grossman2016elasticity, blossey17}  The key predictions of the model can be divided into two-regimes:  narrow- and wide-ribbon regimes.  Narrow-ribbons largely maintain their preferred negative Gaussian curvature, and therefor incur elastic penalties (per unit area) for crystal strains that grow with ribbon width $w$ as $\sim w^4 \kappa_0^4$ where $\kappa_0$ is the preferred curvature radius.  
this super-extensive elastic cost may, in balance with the cohesive drive for larger assembly due to line tension, determine a thermodynamically optimal assembly width that grows with {\it decreasing} curvature, $w_0 \sim \kappa_0^{-4/5}$.
%For narrow-ribbons (low enough line energy) this super-extensive elastic cost balances cohesive drive for larger assembly due to line tension determine an thermodynamically optimal assembly width, $w_0 \sim \kappa_0^{-4/5}$ where $\kappa_0$ is the preferred curvature radius.  

% For large \MJS{wide or long? to match above I think it should be wide.} 
For wide ribbons, the in-plane elastic costs of negative Gaussian curvature overwhelm the (bending) cost to deform ribbons to an unfrustrated shape, leading to a shape transition from helicoids to spiral ribbons which expel Gaussian curvature.  This transition, which we refer to as {\it shape-flattening} throughout this article, is predicted to occur at critical width,  $w_* \sim (B /Y)^{-1/4} \kappa_0^{-1/2}$ where $B$ and $Y$ are respective bending and in-plane (2D Young's) moduli for membrane.  Notably this same underlying mechanical transition has realized in a range of fabricated ribbon architectures, \cite{Armon2014, jeon2017reconfigurable} is proposed as the basis of similar morphological transitions in a range of elastic structures in biology, ~\cite{armon2011geometry, wan2018sensor} and has been verified by a range of finite-element or finite-difference numerical simulations. \cite{selinger2004shape, Armon2014}  For the context of self-assembling ribbons, which can adjust their widths via addition of free subunits, the helicoid-to-spiral transition marks a shape-flattening transition, in which the shape progressively expels Guassian curvature with increasing width. Since the elastic energy becomes extensive in size for $w > w_*$, frustration cannot limit the assembly size in wide ribbon regime, and $w_0 \approx w_*$ marks an {\it upper limit} to the possible range of frustration limited assembly.  For helicoidal assemblies, it is observed that optimal assemblies instead {\it close upon themselves} into finite-diameter cylinders, a.k.a. tubules. \cite{oda1999tuning, selinger96, ziserman2011curvature}

%One current challenge is to connect the continuum description to the characterization of the discrete building blocks \cite{Zhang2019, serafin2021frustrated}. Recent progress in the synthesis of colloidal-scale particles with programmed shape can allow for tunable shape frustration that can more fully test the continuum theory description of size control \cite{hueckel2021total}. Furthermore, advances in DNA nanotechnology allow for both tuning the shape frustration as well as new opportunities for programming the interactions to separately tune the strength of cohesion and costs associated with distinct modes of assembly deformation \cite{sigl2021programmable}. Building from the existing theoretical framework, we consider the effects of engineering such modes as described below.

\begin{figure}
\centering\includegraphics[width=1.0\linewidth]{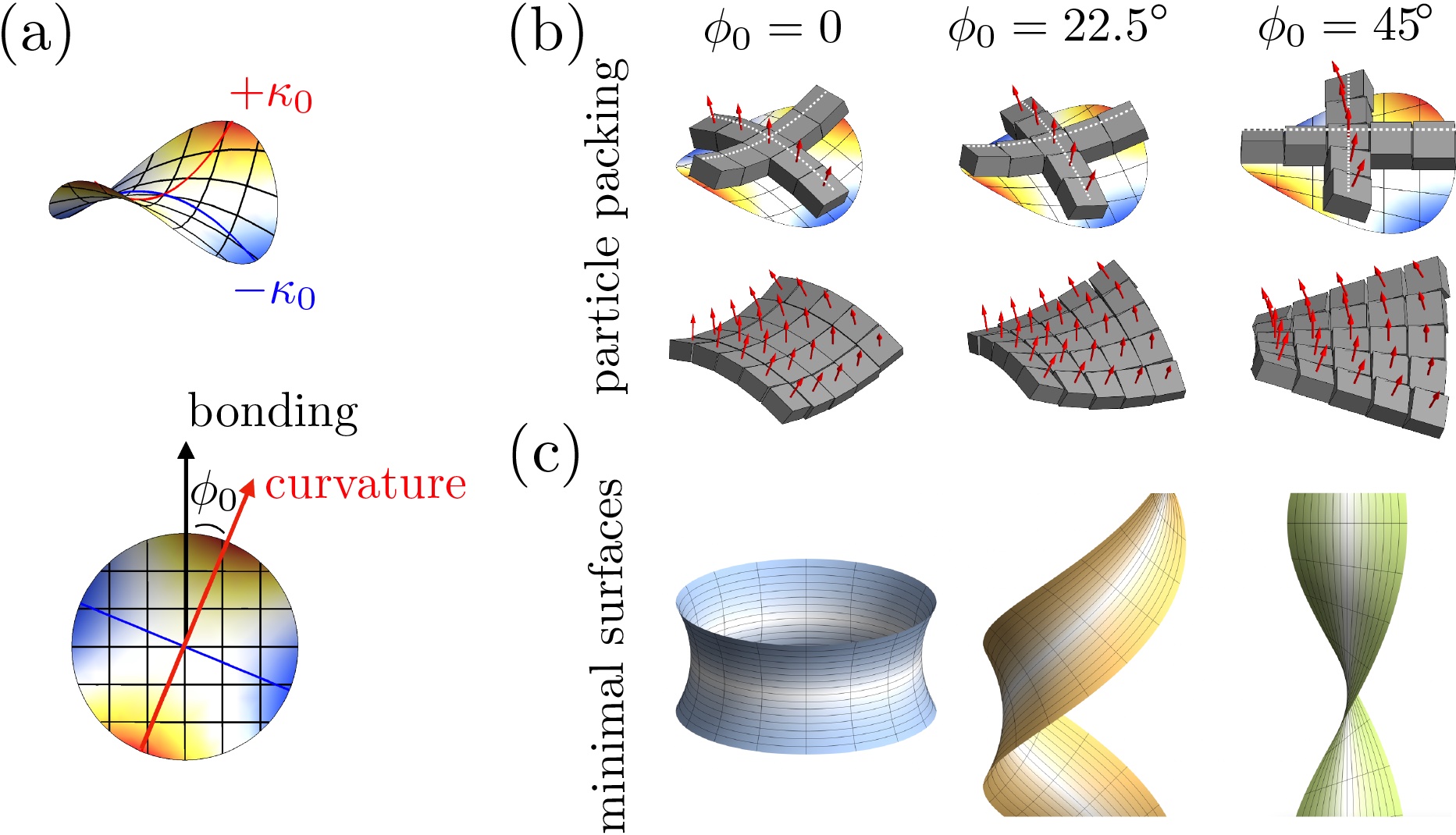} 
\caption{ 
Assembly geomety and the role of $\phi_0$ are illustrated. In (a) the local geometry is a saddle shape with principal curvature $\pm \kappa_0$. Additionally, the principal curvature directions are oriented with respect to the assembly's close-packing directions by an angle $\phi_0$. (b) Particles packed according to this geometry adopt angular differences with respect to their neighbors, rotating parallel to the bonding directions when $\phi_0 =0$ and twisting orthogonal to the bonding directions for $\phi_0 = 45^\circ$. (c) When the assembly forms a ribbon with edges oriented along the close-packed directions, the assemblies are described by the isometric family of surfaces spanning between the catenoid and helicoid.
}
\label{fig:fig_phi_bonnet}
\end{figure}

In this study, we aim to extend the understanding for frustration in hyperbolic, crystalline membranes from the level of phenomenological continuum descriptions to scale of shape-frustrated subunits from they form.  Specifically, we introduce a new class of ``saddle-wedge monomers'' (SWMs) whose large scale interactions favor mesoscopic assembly geometries that map onto the existing continuum models.  As shown schematically in Fig. \ref{fig:fig_phi_bonnet}, the variable shape of the SWM model encodes both tunable frustration (i.e. variable $\kappa_0$) and also programmable relative directions of curvature with respect to close-packed crystalline (bonding) directions.  This broader class of geometries has been recognized as equivalently frustrated, \cite{armon2011geometry} and are related to the isometric (Bonnet) family of minimal surfaces spanning the catenoid and helicoid, \cite{lidin1990some, bonnet1853} parameterized by {\it curvature angle} $\phi_0$ between the crystalline rows and preferred curvature axis.

The aim of this discrete monomer study is several fold.  First, we aim to understand how features of the geometry and interactions of building blocks govern the mesoscale shape and thermodynamics of optimal assemblies, and more specifically, determine the mapping of particle-scale properties onto parameters of the continuum description.  Based on this, we analyze how the range of accessible self-limiting widths compare to size of the building blocks themselves.  We show that the maximal size range of self-limitation is critically delimited by the {\it range} of cohesive bonds between monomers.  Second, we analyze the responses to frustration that fall outside of linear elastic descriptions, more specifically the distinct roles of strain softening and yielding on size control, and the possibility of partial or incomplete bonding.  We show that the former slightly depresses the range of thermodynamic self-limitation, relative to a purely Hookean elastic behavior, while the latter may be associated with a range of heirarchical ground states possible for sufficiently low temperatures.

The remainder of this manuscript is organized as follows. We first summarize a continuum scale description that is expected to capture the mesoscale structure and thermodynamics of SWM assembly, and then introduce the coarse-grained, discrete particle model. Next, we present numerical results from energy minimization calculations on the stress accumulation and flattening of energetic ground states of varying width. We then analyze the limiting case of the flattened state, the tube morphology, with numerics and continuum results considering the effects of anisotropic bending stiffness and strain softening, which are then used to construct the self-limitation phase diagrams in terms of SWM geometry and interactions. Next, we analyze deviations from purely linear-elastic behavior exhibited by the discrete subunit assembly, in particular, show that finite-range interactions generically imply the stability of internally-cracked or weakly-aggregated finite-domain morphologies in regimes where self-limited structures are favored over unlimited (bulk) structures.  Finally, we conclude by discussing the relevance of the results to assembly at finite temperatures, implications for hierarchical assembly and we present preliminary evidence of assembly with molecular dynamics (MD) results.

\section{Models of frustrated hyperbolic ribbons}
We first summarize the key ingredients and predictions of a continuum elastic model for assembly of hyperbolic, crystalline membranes followed by the introduction of discrete particle model of SWM whose assembly forms these frustrated morphologies.

\label{sec: models}

\subsection{Continuum Theory}

\label{sec: continuum}

Here we summarize a continuum elastic description for the frustrated ribbons formed by 2D crystalline membranes with a preference for negative Gaussian curvature shapes (see \ref{section:si_continuum} for full details).  The model, which we refer to as ``narrow ribbon'' (NR) theory, is essentially an elaboration of the original approach of Ref.~\cite{Ghafouri2005}, generalized to include arbitrary direction or curvature axes relative to crystallographic axis, as in Ref.~\cite{Armon2014}.  The approach assumes slender assemblies with an assembly length $L$ and width $w$ such that $L \gg w$: either ribbons of width $w$ much smaller than the (unlimited) assembly length $L$ in the orthogonal direction, or instead closed rings with width $w$ much smaller than the assembly circumference $L \gg w$. The model includes three ingredients,
\begin{equation} \label{eq:model_energy}
    E_{\rm tot} =  E_{\rm bend} + E_{\rm strain}  + E_{\rm edge} 
\end{equation}
corresponding, respectively, to  elasticity of extrinsic (i.e. bending) curvature, in-plane elastic strains of the 2D crystalline order, and the cohesive cost of free edges of the ribbons, dominated by the two longer edges (i.e. $ E_{\rm edge} \sim 2 \gamma L$).  We consider the case of in-plane square-lattice order, and due to the energetics of strong (nearest neighbor) bonding along the lattice directions, assume that optimal ribbons form with their free edges along the lattice directions (i.e. either the local $\hat{x}$ or $\hat{y}$ direction of the ribbons, which are the low edge energy directions).  Here we take the $\hat{y}$ direction to be the long axis of the ribbon).  

General considerations of the elasticity of anisotropic membranes, ~\cite{helfrich1988intrinsic} imply a coupling between free energy to the curvature tensor curvature tensor $C_{ij}$ of the membrane.  According to the narrow-ribbon approximation, for which $|C_{ij}| w \ll1$, we assume that curvatures are roughly constant across the width of membrane, and described by the values at the mid-line: $C_{yy}$ along the ribbon's length, $C_{xx}$ along the width, and $C_{xy}=C_{yx}$ the off-diagonal element of the curvature tensor.  Specifically our systems are described by the following extrinsic curvature elasticity,
\begin{equation}
\label{eq: bend}
E_{\rm bend} \simeq  \frac{w L}{2}  \Big( C_{ij} -(C_0)_{ij}\Big) B_{ijkl } \Big( C_{kl} -(C_0)_{kl}\Big) .
\end{equation}
where $B_{ijkl}$ is the tensor of elastic bending constants and $(C_0)_{ij}$ is the locally preferred curvature. The nearest neighbor binding square-lattice model leads to two non-zero elasticity constants,
\begin{equation}
B_\parallel = B_{xxxx}= B_{yyyy} ; \ B_\perp = B_{xyxy}=B_{yxyx} 
\end{equation}
for deformations that alter bending and twisting (of the tangent plane) along lattice rows; the remaining elastic constants are zero.  The preferred (or ``target'') shape can be written in matrix form,
\begin{equation}
\label{eq: C0}
    {\bf C}_0 = \kappa_0   \left( {\begin{array}{rr}
\cos (2 \phi_0)  & \sin (2 \phi_0) \\
\sin (2 \phi_0) & -\cos (2 \phi_0) \\
\end{array} } \right) ,
\end{equation}
where $\kappa_0$ sets the magnitude of the preferred principle curvatures and $\phi_0$ parameterizes the angle between the lattice directions and the principle curvature directions (see Fig. \ref{fig:fig_phi_bonnet}).  Notably, this preferred curvature targets {\it minimal surfaces} with a mean curvature $H_0 \equiv {\rm Tr} [{\bf C}_0]/2 =0$ and a preferred negative Gaussian curvature $K_{G,0} \equiv {\rm det} [{\bf C}_0] = - \kappa_0^2$.  While the original NR apporach of Ghafouri and Bruinsma \cite{Ghafouri2005} for chiral membranes corresponds to the case of $\phi_0 = \pi/4$, it was pointed out by Armon and coworkers that a larger family of target minimal surfaces (corresponding to the Bonnet family of minimal ribbons) are generated simply by rotation of the preferred curvature axis relative to its pitch axis.~\cite{Armon2014}  We explore the implications of this broader control over frustrated shape for the design of the SWM and its ultimate assembly below.  

The strain elastic energy takes the from 
\begin{equation}
    E_{\rm strain} = \frac{1}{2} \int dA ~ \sigma_{ij} u_{ij}
\end{equation}
where $u_{ij}$ and $\sigma_{ij}=\lambda u_{kk} \delta_{ij} + 2 \mu u_{ij} + \lambda_\perp \big( \delta_{ix}\delta_{jx} u_{yy} + \delta_{iy}\delta_{jy} u_{xx}\big) $ are the in-plane 2D strain and stress tensors for a square crystal.  In-plane strains are coupled to the out-of-plane deflection the membrane through its intrinsic curvature, i.e. non-zero Gaussian curvature generates in-plane stress gradients.~\cite{Seung1988}  As described in Ref. ~\cite{Ghafouri2005} and in the Appendix, these may be solved for long-ribbons assuming uniform stress along $y$ and constant Gaussian curvature $K_G$ resulting in an elastic cost that grows superextensively with width yielding,
\begin{equation}
\label{eq: strain}
E_{\rm strain}/A \simeq \frac{Y}{1440} K_G^2 w^4 ,
\end{equation}
where $Y=(2\mu - \lambda_\perp)(2 \lambda + 2 \mu + \lambda_\perp)/(\lambda + 2 \mu)$ is the 2D Young's modulus of the membrane.

\begin{figure}
\centering\includegraphics[width=0.9\linewidth]{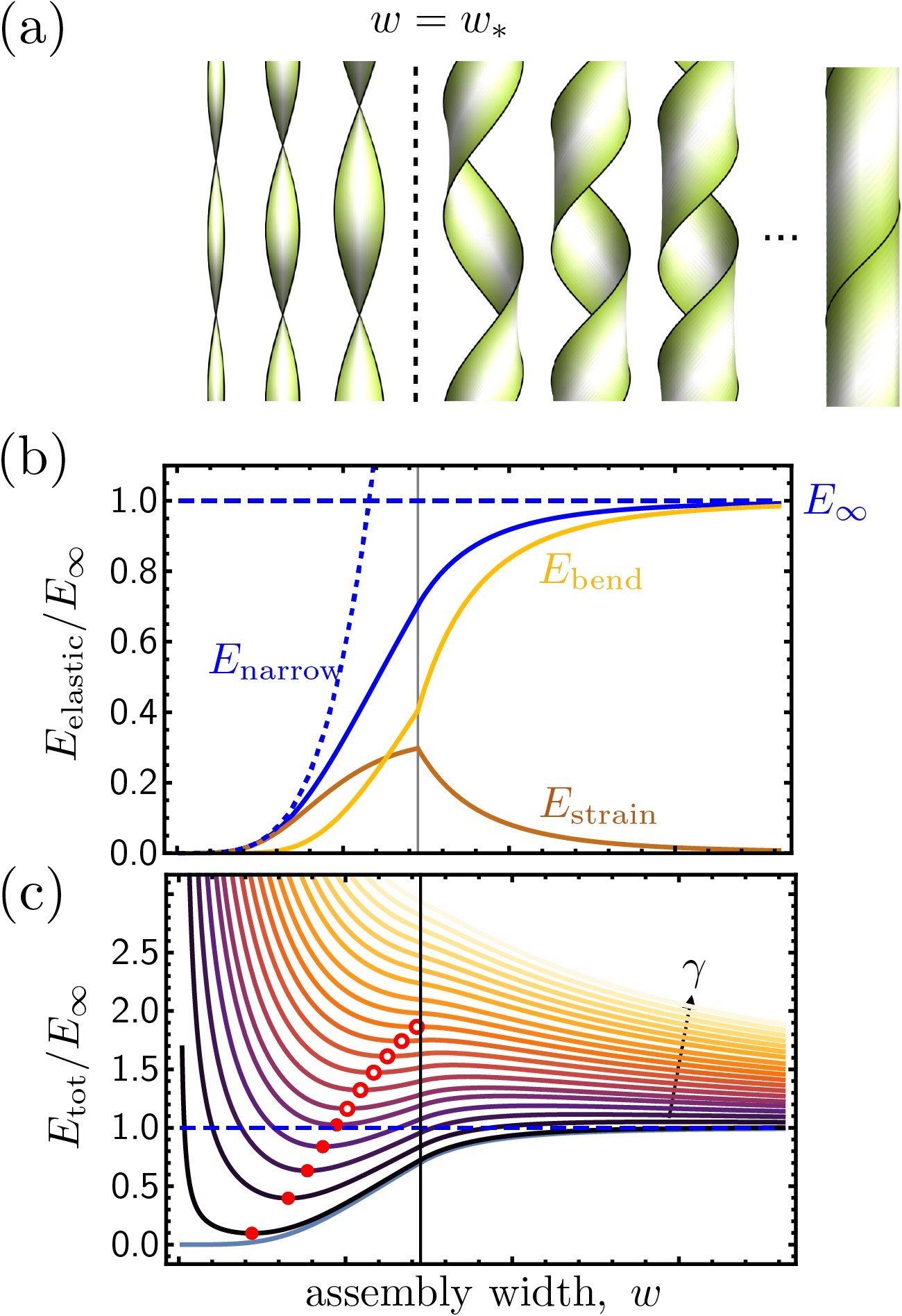} 
\caption{ 
(a) Ribbons of increasing width $w$, for the case $\phi_0=45^\circ$, maintain approximately helicoidal shape up until a width $w_*$ at which point they flatten to a cylindrical shape.  The flattened ribbons may continue widening until the edges meet and a closed tubule is formed. 
(b) Schematic plots of $E_{\rm{bend}}$ (yellow), $E_{\rm{strain}}$ (brown) and  $E_{\rm{elastic}} = E_{\rm{bend}}+E_{\rm{strain}}$ (solid blue curve), as a function of the assembly width $w$. Assymptotic limits are shown for narrow ribbons $E_{\rm{elastic}}(w\to 0) = E_{\rm narrow} \propto Y \kappa_0^4 w^4$ (dotted blue) and wide ribbons $E_{\rm{elastic}}(w\to \infty)=E_\infty$ (dashed blue).
(c) A self-limiting state is a minimum in the model free energy $E_{\rm tot} = E_{\rm elastic} + E_{\rm edge}$. With increasing line tension $\gamma$, larger self-limiting sizes may be achieved up until a point when the minimum's energy is equal to the flattened state's energy $E_\infty$ at which point the self-limiting state is metastable. Further increase of the line tension results in larger metastable finite minima until the point at which the minimum disappears entirely near $w=w_*$.}
\label{fig:fig_gb_sla_theory}
\end{figure}

The thermodynamics of the NR approximation follow from minimization of the total free energy density with respect to curvature and ribbon width and are summarized schematically in Fig.~\ref{fig:fig_gb_sla_theory}.  For narrow ribbons (corresponding to small $\gamma$), the ribbon adopts a shape close it is target hyperbolic shape, ${\bf C}(w \to 0) \simeq {\bf C}_0$, so that the dominant elastic costs derive from in-plane strains.  As a result, the stretching energy is super-extensive, growing faster than the assembly size $A$, according to $E_{\rm strain}/A \sim Y \kappa_0^4 w^4 $. In this regime, the optimal width $w_0$ is set (approximately) by the balance between in-plane stretching of the target shape and the edge energy (per unit area), $E_{\rm edge}/A \sim \gamma/w$, leading to an optimal (self-limiting) width that grows with edge energy and decreases with increasing target curvature, $w_0 \sim (\gamma / Y \kappa_0^4)^{1/5}$.  When ribbons grow sufficiently large, the strain energy cost to maintain the preferred negative Gaussian curvature overwhelms the cost to unbend that assembly into an isometric (i.e. $K_G \to 0$) shape.  Roughly speaking this occurs at a characteristic width scale, $w_* \sim (B / (Y \kappa_0^2))^{1/4}$

The model energy functional gives the (approximate) optimal shape $C_{ij}(w)$ and resulting elastic energy $E(w)$ as a function of ribbon width (shown schematically in Fig. \ref{fig:fig_gb_sla_theory}), when optimized over values of the curvatures. For small curvatures,  $E(w)/A \simeq  Y \kappa_0^4 w^4 / 1440$ as the curvature remains close to the preferred value.  When ribbons reach a critical width $w_*$, the strain energy cost to maintain the preferred negative Gaussian curvature overwhelms the cost to unbend that assembly to reduce $K_G$, and the elastic ground states undergo a symmetry breaking bifurcation.  For chiral ribbons ($\phi_0 = \pi/4)$, this shape transition corresponds to a transformation from helicoids to spirals.  In ~\ref{section:si_continuum} we show that the supercritical shape transition occurs in GB theory for any value $\phi_0$ and for $0<\phi_0<\pi/4$, leading to two different spiral equilibria (degenerate within GB theory). See for example the two stable branches for $w > w_*$ for $\phi_0=22.5^\circ$ in Fig.~\ref{fig:fig_flattened_branches}, which different in terms of helical pitch and radius.  

In the limit $w \to \infty$, the the Gaussian curvature vanishes at the expense of bending spiral membranes into cylindrical shape with energy,
\begin{eqnarray}
\frac{E_\infty}{A} = \frac{1}{2}B_{\parallel}\kappa_0^2 \left[1 + \frac{B_\perp - B_\parallel}{B_\perp + B_\parallel} \sin^2(2\phi_0)\right]. \label{eq:gb_einfty}
\end{eqnarray}
The flattened cylindrical shape is identified with frustration escape, as the assembly can grow without increasing elastic energy density. At large enough widths, the actual assembly will close up so that the flattened state is a closed tubule. Whereas the initial stretching cost is independent of $\phi_0$, the expression for the flattening cost $E_\infty$ may depend on $\phi_0$ when the associated curvature moduli differ $B_\perp \neq B_\parallel$. One might naively expect that the larger flattening cost can extend the range of super-extensive elastic energy with growing size, and thus increase the range of size control; that is, the mechanical equilibrium would shift from self-limiting ribbon shapes to flattened, tube morphology roughly speaking when the ribbon elastic energy $E/A \simeq  Y \kappa_0^4 w^ 4/1440 $ was equal to $E_\infty/A$, so that $w_\mathrm{max} \sim (E_\infty/Y \kappa_0^4)^{1/4}\approx w_*$ would increase with increasing flattening cost.  That is, based on this model, the mechanics of unbending the membrane away from its curved shape sets an upper limit to size scales where frustration can provide a thermodynamic limitation to the ribbon width.  Analysis of equation \ref{eq:model_energy} predicts a moderate reduction in the range of self-limitation with increasing curvature angle, $\phi_0$, as a consequence of the mechanical flattening transition occurring at a \textit{smaller} value of $w_*$ with increasing flattening cost $E_\infty$. The central goal of this study is to directly assess variation of the range of frustration-limiting widths with $\phi_0$, as the target shape is varied from catenoidal to helicoidal, for a discrete subunit model of hyperbolic, 2D crystalline membrane assemblies.

We note that the assumptions of the NR theory, namely that curvatures are sufficiently uniform across the width of ribbons, do not strictly hold across the full range of ribbons widths.  This is because torque-free boundary conditions require a boundary layer of characteristic size proportional to $w_*\sim (B/Y)^{1/4}\kappa_0^{-1/2}$, \cite{efrati2009buckling, Armon2014, arieli2021geometric} so that through-width curvature variation becomes non-negligible for $w > w_*$.   We show this for an explicit solution for exact (boundary layer) solution below (for $\phi_0 =0$ in Appendix ~\ref{section:si_continuum}). This boundary layer correction modifies predictions of elastic energy, particularly in the large $w \gtrsim w_*$ regime.  Notably, finite-element calculations for ribbons $0<\phi_0<\pi/4$ suggest that boundary-layer corrections break the degeneracy between the two large-$w$ equilibria. These detailed corrections for intermediate-$w$ notwithstanding, we argue that the NR approximation works reasonably well for both the self-limiting (i.e. $w$ small) and the asymptotically flattened ($w \to \infty$) regimes.  The simple and analytically tractable solutions of the NR theory therefore provide a useful means to survey how thermodynamics of self-limitation varied with geometric and mechanical properties of the membrane in the continuum elastic description.

%Corrections to the theory derived in \cite{efrati2009buckling} and referenced in the appendix, suggest a competing effect that promotes increasing $w_\rm{max}$ with $\phi_0$, due to the extrinsic geometry and mechanics at the boundary. A central result presented in this study is to test the range of size limitation in the discrete wedge-monomer model.  

\begin{figure*}[h!]
% \centering\includegraphics[width=1.0\linewidth]{fig_monomer_design.jpg} 
\centering\includegraphics[width=0.7\linewidth]{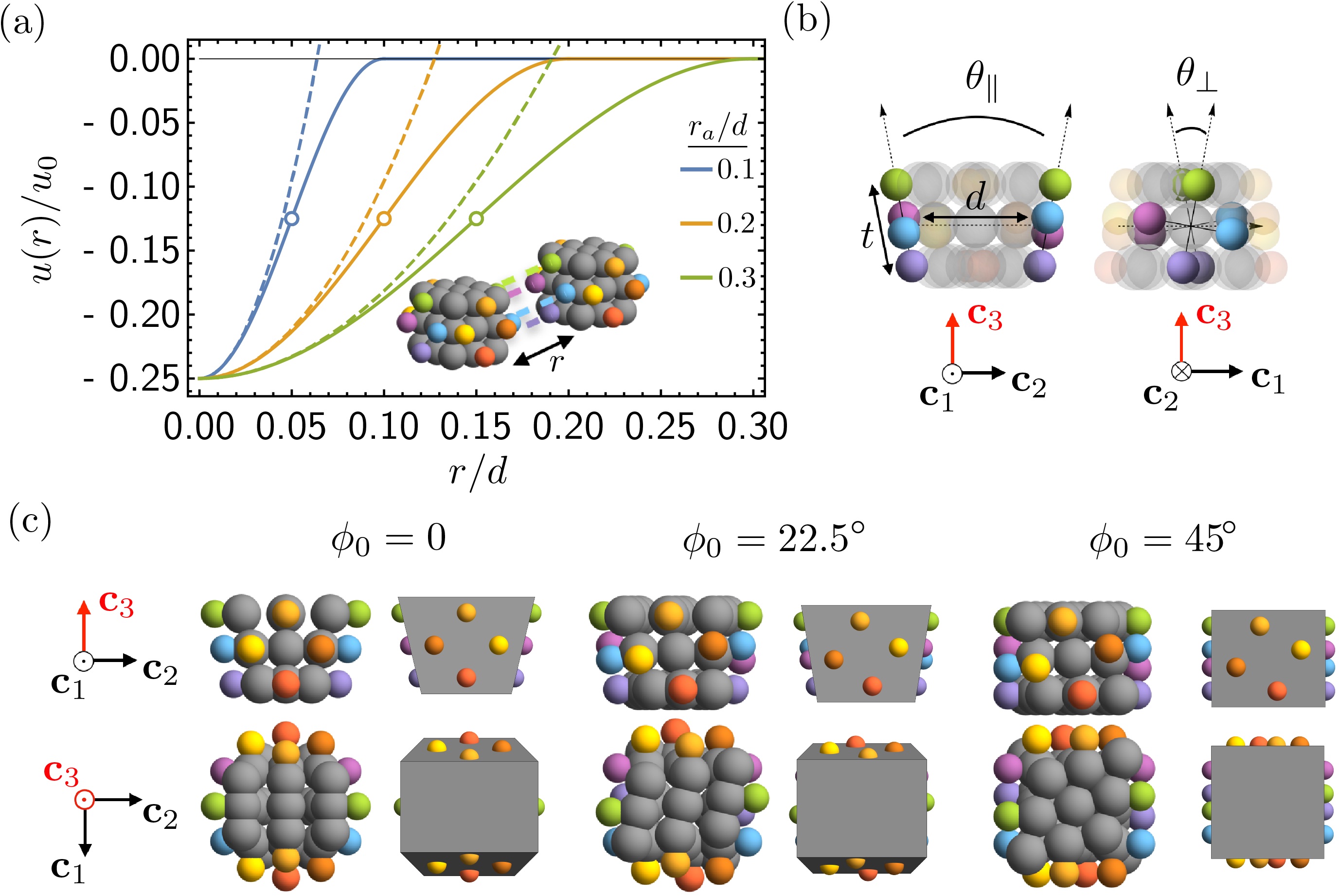} 
\caption{ Monomer design for programming assembly size control. (a) The attractive site pair potential $u(r)$ is plotted as a function or separation $r$ for varying potential range $r_a$ with respect to monomer size $d$ and bond energy $u_0$. The yield point for each case is indicated with an open circle and the harmonic approximation plotted as a dashed curve. (b) A sphere representation of the monomer design viewed along the ${\bf c}_1$ and ${\bf c}_2$ monomer bonding axes. Gray spheres have radius representing the range of their excluded volume interaction. Attractive sites are spheres of different colors, with four attractive sites on each of the four bonding faces of the monomer. The four attractive sites on a given side define a frame with attractive site spacing $t$ between opposite pairs. The frame between opposite sides has distance $d$ defining the monomer size, and rotation $\theta_\parallel$ projected along the frame displacement and $\theta_\perp$ orthogonal to the displacement. (c) 
% The preferred assembly geometry is a surface with principal curvatures $\kappa_1 = + \kappa_0$ and  $\kappa_2 = - \kappa_0$ and bonding directions making angle $\phi_0$ with respect to the principal curvature directions. \MJS{There is no part (d)} 
Monomer geometries for $\phi_0 = 0,22.5,45^\circ$  are shown viewed along the ${\bf c}_1$ bonding axis and along the $c_3$ axis (red) orthogonal to the assembly surface. Sphere representations are shown alongside respresentations with the excluded volume of each monomer represented as a polygon.
}
\label{fig:monomer_design}
\end{figure*}

\subsubsection{Discrete Model and Methods}
\label{sec:model_methods}
To connect discrete-monomer properties and design to assembly behavior, we developed a coarse-grained simulation model, building from a model previously developed to study microtubule assembly. \cite{cheng2012self, cheng2014self, stevens2017long,Bollinger2018, bollinger2019diverse} The saddle-wedge monomer (SWM) is designed for energy minimization and dynamical assembly simulation using the LAMMPS software. \cite{plimpton1995fast,Thompson2021, linklammps}  The basic shape of the SWM is a ``double-wedge'' geometry: four binding faces that promote curvature of opposite signs in the orthogonal directions of the assembly, as illustrated in Fig. \ref{fig:monomer_design}. The rigid monomer consists of 27 sites of a single type and purely repulsive interactions, surrounded by 16 attractive sites of 8 types, with the attractive sites arranged in a planar square on each of four bonding sides of the monomer. The square diagonal of the attractive sites on each face defines the {\it thickness} parameter $t$ that in principle may be used to tune the relative costs of changes in assembly curvature, bending, with respect to assembly stretching. 
% \MJS{This sentence is not used, at least here or is this what is in Fig. 3?: In the results presented below consider the case $t\simeq 0.568 d$.} \DMH{replaced with the following:}
For the results presented in this study, we consider the case $t = 0.568 d$.
The monomer width $d$ is defined by the distance between respective centers of mass of attractive sites on opposite sides. 
For the attractive sites, pairwise binding only acts between sites of the same type on different monomers (as denoted by distinct colors of binding sites in Fig. \ref{fig:monomer_design}). The repulsive sites interact according to a Weeks-Chandler-Anderson (WCA) pair potential, \cite{weeks1971role} and define the monomer excluded volume and shape in the low-energy minimized structures. Their arrangement, with coordinates, are described in more detail in Appendix  \ref{section:si_monomer_design}. 
%MJS I changed the next sentence
%DMH I see the improvement, but am worried about 'plane of repulsive sites' since the 9 repulsive sites are not a plane for phi>0 ... I will think how to keep getting more concise regarding repulsive sites, especially in the main text. 
The attractive sites on a given face are arranged in a plane parallel to the adjacent plane of repulsive sites on the monomer, so that pairwise attractions of all four sites are possible without overlaps from purely repulsive sites.
%The repulsive sites are arranged so that contact between all repulsive on two faces requires the same monomer orientations (respective angles) as contact between all attractive sites (see appendix). 
The attractive site interactions each have the form of 
\begin{eqnarray}
u(r)= \left\{
 \begin{array}{ll}
 \displaystyle -\frac{1}{8}u_0\left[1 + \cos \left( \frac{\pi r}{r_a}\right) \right], &   r \leq r_a\\
 0, &  r > r_a \\
\end{array} \right. ,
\label{eq:attractive}
\end{eqnarray}
where $r$ is the distance between interacting sites, $r_a$ is the interaction range and $u_0$ defines the potential well depth, such that the minimum energy for two monomers binding with all four attractive sites ideally placed is $-u_0$.
% is the (unstrained) bond energy and . %MJS This is true, but gives a simulation person pause about whether u_0 is a constant.
Figure \ref{fig:monomer_design}(a) shows the shape of the attractive interaction with varying interaction range. Full details of the monomer geometry and interactions are given in Appendix \ref{section:si_monomer_design}. Importantly, $r_a/d$, the range of interaction with respect to the monomer width, controls both the relative stiffness of the assembly via the elastic moduli defined below and also the strain necessary for a single bond to reach the point of yielding.

The orientational geometry %of bound SWM 
relating pairs of SWMs bound together is defined in terms of the orthonormal frame $\{{\bf c}_1,{\bf c}_2,{\bf c}_3\}$ associated with each monomer as shown in %Fig. \ref{section:si_monomer_design}
Fig. \ref{fig:monomer_design}(b)
with the first two directions pointing along neighbor bonding axes and the third direction point along the vertical (non-bonding) direction. The preferred binding geometry is determined by angles defined in a single rigid monomer.   Attractive sites are arranged so that $ \theta_0 $ is the preferred angle between the ${\bf c}_3$ axis of bonded neighboring monomers, when all four interacting sites on their respective faces coincide.  The $\phi_0$ angle can be understood as a twist of binding directions (i.e. orientation of the square of attractive sites) around the axes connecting the SWM centers to their binding faces, i.e. by + or - around ${\bf c}_1$ and ${\bf c}_2$, respectively (the twist sense in one bonding direction is chosen to be opposite that in the other direction to be compatible with membrane geometry of zero mean curvature).   Taking the ${\bf c}_3$ direction to be normal to the mid-surface of %for
multi-particle membrane assemblies formed by SWMs, we can relate wedge angle $\theta_0$ and curvature angle $\phi_0$ to the target curvature tensor of the membrane as follows.  Projecting the  rotation sense ${\bf c}_3$ between ${\bf c}_i$ faces into the ${\bf c}_j$ gives the preferred surface curvature $C_{ij}$ times the particle width $d$, or the angles $\theta_{\parallel} = \theta_0 \cos 2 \phi_0$ and $\theta_{\perp} = \theta_0 \sin 2 \phi_0$ as illustrated in Fig. \ref{fig:monomer_design}(b).   Hence, orientational geometry of SWMs map onto preferred curvature of the form eq. (\ref{eq: C0}) with target principle curvature 
\begin{equation}
    \kappa_0 = \theta_0/d .
\end{equation} 
The attractive site arrangement defines both a monomer width $d$ that is approximately the preferred distance between neighboring monomers and a monomer thickness $t = 0.568 d$ that controls the cost of bending deformations. As described in Appendix \ref{section:si_continuum}, the effective elastic constants of membrane assemblies of SWM are determined by consideration of the local deformations on ideally bounded neighbors, imposed by distortions of a crystalline membrane.  As shown schematically in Fig. \ref{fig:fig1a}, $Y$ corresponds to stretching/compressing of inter-face spacing, while $B_\parallel$ and $B_\perp$ correspond to dihedral and twist angle distortions between bound SWM.  Modeling bound attractive sites as effective springs of stiffness $\pi^2 u_0/r_a^2$ leads to
\begin{eqnarray}
\label{eq: params}
Y = \frac{\pi^2 u_0}{2 r_a^2}; \  B_{\parallel} = \frac{\pi^2 t^2 u_0}{16 r_a^2}; \ B_{\perp} = \frac{\pi^2 t^2 u_0}{8 r_a^2}.
\end{eqnarray}
Notably, as SWM are modeled as rigid bodies, eq. (\ref{eq: params}) highlights the role of the {\it range} of attraction in controlling the deformability of the assembly.  Additionally, we note that the characteristic ratio of bend to stretch moduli $B/Y\propto t^2$ is {\it independent} of interaction parameters, controlled only by the geometric thickness of the SWM particles.  Last, it is important to note that, distinct from previously studied models of anisotropic bend-elasticity, off-diagonal bending is stiffer that bending along the lattice directions (i.e. $B_{\perp}=2 B_{\parallel}$. The greater twist stiffness relative to row bending, is generic consequence of attraction only binding, and has consequences in the thermodynamics of frustration escape for distinct $\phi_0$ values of SWM. As discussed further below, one consequence is the dependence of the flattening transition on $\phi_0$, 
\begin{equation} \label{eq:discrete_swm_wstar}
w_* =(360)^{1/4} \frac{\sqrt{t /  \kappa_0}}{ \sqrt{1 + \frac{7}{9}\sin^2(2\phi_0)}}  =3.28~ \frac{ d / \sqrt{\theta_0}}{ \sqrt{1 + \frac{7}{9}\sin^2(2\phi_0)}} ,
\end{equation}
which is derived in Appendix \ref{section:si_continuum}, eqs. (\ref{eq: elastic})-(\ref{eq: wstar}).  As we consider assemblies to form with open edges only along {\it low energy} directions in the bond lattice, it is straightforward to compute the edge energy per unit length 
\begin{equation}
    \gamma = \frac{u_0}{2 d}, 
\end{equation}    
as (half) the ideal bond energy needed to separate membranes along their nearest neighbor direction.

To explore the the groundstate thermodynamics of size control of this model, the LAMMPS simulation software was used to minimize the energy of preassembled initial configurations. The LAMMPS \textit{minimize} command was used with default, conjugate gradient method. Structures were successively minimized from a soft, relatively long-range interaction $r_a/d = 0.199$ down to the target range of interaction,
% \MJS{ should it be 'decrementing'?}
decrementing $r_a/d $ by $0.007$ and re-minimizing at each step. Additionally, the monomers were first minimized at a softer state where intra-monomer geometry was maintained with springs: (see appendix) minimizations with incrementing $r_a$ were run at intra-monomer bond stiffness $k_\mathrm{bond} = 885 ~ u_0 / d^2$ %1000
and subsequently the minimization at the final value of $r_a$ was re-minimized at successive values  $k_\mathrm{bond}/( u_0 / d^2) = 890, 4400, 8900, 44000$ and $89000$. The intra-monomer bonds are found to contribute negligible total energy compared to the total inter-monomer interaction potential energy (less than 1 part in $10^5$) after minimization at the larger bond stiffness.   
The final minimization was run until reaching a force tolerance of $0.3 \times 10^{-4} u_0/d$. 
%($10^{-4} \epsilon/\sigma$)
Intermediate steps at higher $r_a$ and lower $k_\mathrm{bond}$ were run until either the same force tolerance was reached, or $10^8$ steps of minimization.

%here is defined assembly parameter $w$ :
Multiple pre-assembled initial configurations were sampled with varying lateral dimensions $w/d$ corresponding to the ring or ribbon width, which is the number of monomer rows as measured in the shorter assembly direction. The initial configurations for varying width were cylindrical geometry $\phi_0 = 0$ and flat rectangular geometry for $\phi_0 >0$. The flattened tubule state's energy was found from initially cylindrical geometry for longer tubes. For all starting configurations, monomers were arranged at a slightly dilated spacing of $1.05 d$. For $\phi_0 = 0$, the cylinder circumference was chosen to be $360^\circ/\theta_0$, the cylinder length to be the target assembly size of $w/d$ monomers, with monomer bond directions aligned along the circumferential and longitudinal directions. For large-size energetics of all $\phi_0$, cylindrical configurations were prepared with successively varying lengths of  100, 110, 120, 130, 140,  and 150 monomers and bond directions along the cylindrical surface making an angle $\phi_0$ with the tube circumferential direction and axis. For $\phi_0>0$, a rectangular geometry was used to sample smaller assembly size. The bonding directions were chosen in-plane and parallel to the boundaries of the rectangle, one side length was kept to be $100$ monomers while the other chosen to be the target size of $w/d$ monomers.  

To determine optimal zero-temperature size of small-width assemblies, varying cylindrical ring or rectangular ribbon assembly widths $w$ were sampled up to $2 w_*$ where $w_*$ is the theoretical transition width from Eq. \ref{eq:discrete_swm_wstar}. Structures rendered in figures are shown with effective strain energy calculated from the average soft interaction energy, subtracting off the reference value of $-u_0$ for each bond that was present in the initial configuration. Structures were analyzed to determine if any bond initially present in the starting configuration exceeds the yield point in the final relaxed state.

The total assembly energy $U_\mathrm{a}$ is evaluated in terms of all pairwise interactions between sites on different monomers,
\begin{eqnarray}
U_\mathrm{a} = \frac{1}{2}\sum_{i,j} \Big( u(r_{ij}) + u_\mathrm{WCA}(r_{ij}) \Big).
\end{eqnarray}
Following similar analyses of geometrically frustrated assemblies,~\cite{hagan2021} we define the {\it  excess energy} as energy of the assembly relative to the cohesive bulk and edge energetics, and compute it by subtracting the ideal energy $u_0$ (of an unstrained bond) for every bond in the assembly
% Thanks I fixed the factor of two on the line tension! The mistake was in this equation but I have been using the correct expression in my analysis DMH 
% that's counting bonds of rectangular patch:  u_0 \times [2 N_x N_y - N_x - N_y]. 
\begin{eqnarray}
E_\mathrm{ex} = U_\mathrm{a} + 2 u_0 w L /d^2 - u_0 (w + L) / d.
\end{eqnarray}
In the following sections, we consider the comparison of the excess energy of the discrete model $E_\mathrm{ex}$ to the continuum model predictions derived from equation \ref{eq:model_energy}.

\section{Results}
\subsection{Stress accumulation and flattening, beyond harmonic and isotropic elasticity}

In this section, we compare simulated ground states of discrete-SWM assembly to the predictions of the continuum theory, illustrating how strain accumulation and elastic shape-flattening depend on on arrangement of attractive sites and the tapered shapes SWM binding. %The finite range of the interaction results in strain softening, most clearly seen in the results on the flattening energy presented here.

% MJS moved figure to get on p. 7
\begin{figure*}
\centering
\includegraphics[width=0.8\linewidth]{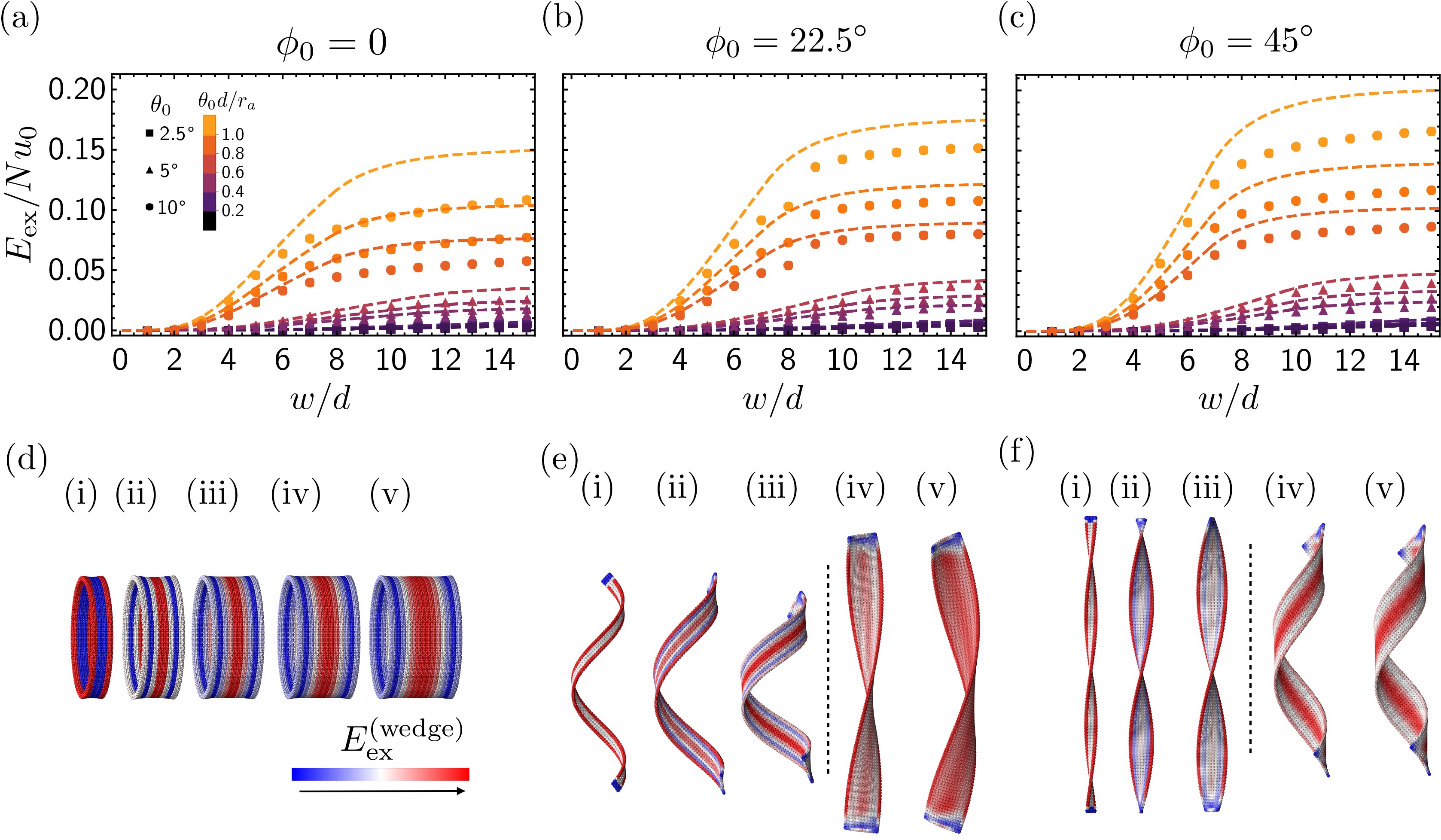}
\caption{ 
Excess energy accumulation and flattening for assemblies of increasing width, for $\phi_0 = 0, 22.5, 45^\circ$, $\theta_0 = 2.5, 5, 10^\circ$ and $r_a/d = 0.14, 0.17, 0.20$. (a-c) Average excess energy density $E_{\rm ex}/Nu_0$ after numerical minimization (points), for varying assembly width $w/d$ and monomer parameters $\theta_0,r_a$, is compared to the continuum theory $E/Nu_0$ prediction (dashed curves).
(d-e) The progression of structures with increasing width generically shows a transition to flattened shape, close to a critical width $w_*$ defined in Eq. \ref{eq:discrete_swm_wstar}.
% from predicted continuum theory $w_*$ \GMG{we need to quote what formula we are using for $w_*$ in the main text if we are going to be referring to comparisons to this prediction}
Structures have widths $w/d = 4, 8, 10, 12, 14$, $r_a/d=0.14$ and $\theta_0 = 5^\circ$. Monomers are colored by relative excess energy $E^{({\rm wedge})}_{\rm ex}$. 
% \DMH{ my attempt at resampling did not improve things}
} \label{fig:stress_accumulation}
\end{figure*}

 We focus on three values of curvature direction, with $\phi_0=0$ the closed ring that approximates a catenoid surface in the narrow limit, $\phi_0 = 45^\circ$ approximating a helicoidal ribbon in the narrow limit, and $\phi_0 = 22.5 ^\circ$ an intermediate case. The excess elastic energy $E_{\rm{ex}}$ computed from energy minimizations are plotted along with (narrow-ribbon) continuum model predictions in Fig.   \ref{fig:stress_accumulation}(a-c), for varying taper angle ($\theta_0$), curvature direction ($\phi_0$), and interaction range ($r_a/d$). In each case, the continuum model accurately captures the excess energy in of discrete assemblies in the $w \to 0$ regime as well as the transition from super-extensive growth at small $w$ to extensive growth (i.e. saturated $E_{\rm{ex}}/N \sim w^0$) at large $w$. Typical structures for the progression through the mechanical transition are rendered in Fig.  \ref{fig:stress_accumulation}(d-f).  The cases for $\theta_0\neq0$ both show an apparently sharp shape transition between low- to high-$w$ values, as highlighted by the dashed lines in Fig.  \ref{fig:stress_accumulation}(e-f).  In Fig.~\ref{fig:fig_flattened_branches}, we compare the predicted curvatures of from NR theory to shapes from the SWM ribbon minimizations in Fig.~\ref{fig:fig_flattened_branches} for $\phi_0=22.5^\circ$ and $45^\circ$.  The general $w$-dependence of simulated ribbon shapes is well-captured by the NR model, with abrupt changes in shape occuring near to the predicted values of $w_*$.  We note that our energy minimizations seemed to resolve only the larger pitch solution for large-$w$ cases of the $\phi_0=22.5^\circ$, and attempts to seed and sample the lower-pitch branch were unsuccessful in finding these equilibria.  This, combined with the observation from finite-element simulations~\cite{Armon2014} that the larger pitch branch has {\it higher} elastic energy, would seem to account for an apparent jump in the computed $E_{\rm ex}$ value for $\phi_0=22.5^\circ$ as the ribbon shape transitions from low- to high-$w$ equilibrium shapes (i.e. visible in between $w/d=8$ and 9 in Fig.~\ref{fig:stress_accumulation}b).  In Fig. ~\ref{fig:fig_gb_sla_theory}a-b, we show, nevertheless, that the basic dependence of $w_*$ on curvature angle $\theta_0$ is well captured by NR theory for both $\phi_0=22.5^\circ$ and $\phi_0=45^\circ$ ribbons.  As there is a no symmetry breaking transition for catnoidal ($\phi_0=0^\circ$) membranes, the exact (boundary layer) formalism summarized in eqs. (\ref{eq: BLcat})-(\ref{eq: wmaxcat}) shows that the shape evolution with increasing $w$ is fully-continuous for this case. No attempt to extract a shape-flattening size from simulated $\phi_0=0^\circ$ membranes was made.
 
 %\GMG{We need to include some explicit statements about where the shape transitions occur in the discrete assemblies and how measure that.} \DMH{Incorporating references to analysis in appendix here}  For the case of $\phi_0 = 22.5^\circ$, numerical results show a slight a jump in the excess energy near to the apparent transition, which is a consequence of the bistable nature of the flattened assembly, with the under-wound shape of larger pitch in structures $(iv)$ and $(v)$ of higher energy than the smaller-pitch structure $(iii)$ 
 % \GMG{I can't tell if we are same that we are not able to resolve the energy of the discrete ground state, or if we beleive the energy density is ACTUALLY discontinuous (DUBIOUS).  Need to be more clear, and probably quickly summarize additional (presumably unsuccessful) efforts to better sample these structures.} 
 %The difference in energy of the two states is not captured in the approximate continuum theory, but was previously shown in numerical results in \cite{Armon2014}. 
 
%MJS moved figure 5 to get on p.9
\begin{figure*}[h!]
% \centering\includegraphics[width=1.0\linewidth]{fig3.jpg}
\centering\includegraphics[width=0.7\linewidth]{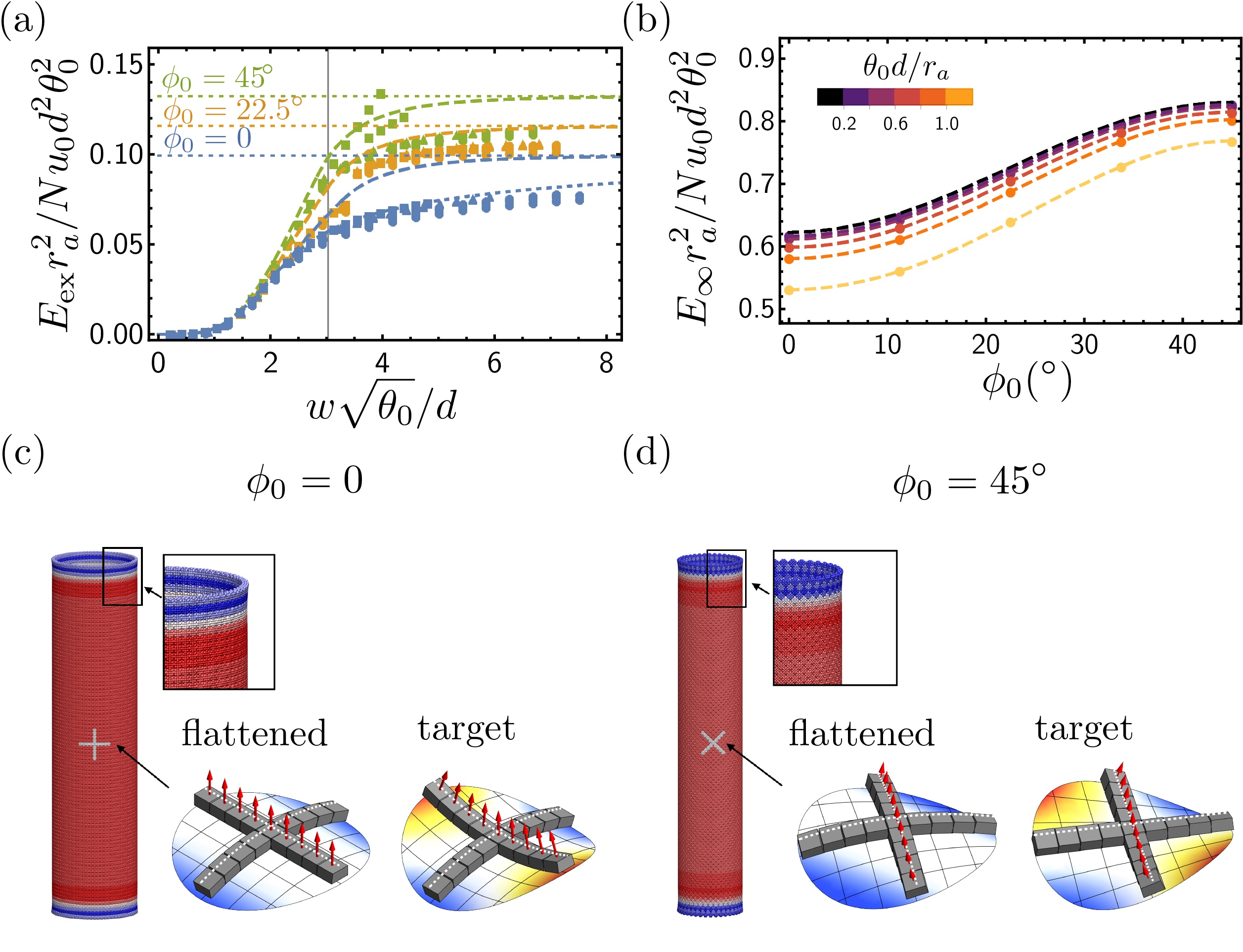}
\caption{ 
Flattening energetics with nonlinear and anisotropic flattening cost. (a) Rescaling the same data (points) in Fig.  2 for $r_a$ and $\theta_0$, shows the collapse and agreement with continuum theory for assembly widths up to the transition, where stress accumulation grows $E_{ex}/N \sim Y \kappa_0^4 w^4$. Dashed curves show the continuum theory for varying $\phi_0$, while the single dotted curve shows a correction in the shape solution for $\phi_0 = 0$ that agrees with the simpler solution (dashed curve) for $w \rightarrow \infty$. Dotted flat lines indicate values of $E_\infty$ from the model prediction for varying $\phi_0$. The vertical line indicates the predicted value for the flattening transition $w_*$. (b)  Extrapolations of numerical minimizations of tubes to infinite length, shown as discrete points, compared to flattened-state solutions of the nonlinear continuum bending energy shown as dashed curves. 
% \MJS{ this is discussion that could be removed from caption: The flattening cost $E_\infty$ increases with $\phi_0$ due to the anisotropic bending stiffness of the assembly deriving from the attractive site arrangements. Furthermore, the finite range of the attractive potential and associated strain-softening results in decreasing flattening cost with increasing $\theta_0/r_a$.}
(c) Minimized tubule assemblies of length $w/d=100$, optimal radius and bond orientations, illustrating the fully flattened geometry of wedge assembly for differing $\phi_0$, away from the assembly boundaries. For $\phi_0=0$, bonds are aligned parallel to the tube axis upon flattening and parallel to the principal directions in the ideal `target' geometry. For $\phi_0=45^\circ$, bond directions are at $45^\circ$ with respect to the tube axis (target principal directions) for the flattened (target) geometries.
}
\label{fig:flattening}
\end{figure*}

 The comparison of discrete and continuum results for $E_{\rm ex}(w)$ is made clearer when results are rescaled by the parameter combination $ N u_0 d^2 \theta_0^2/r_a^2$ (proportional to shape flattening energy)  and rescaling widths by the characteristic elastic scale $d/\sqrt{\theta_0} \sim w_*$ in Fig.  \ref{fig:flattening}(a). Here, the results show good agreement with the approximate continuum theory, plotted as dashed lines, for the limit of small $w$. In this limit, the curves coincide for varying $\phi_0$, showing the monomers are equivalently frustrated with $E_{\rm a}/A \sim  Y \kappa_0^4 w^ 4/1440$ independent of $\phi_0$. Beyond the flattening transition, a noticable discrepancy is due to the continuum model approximation of uniform curvature, whereas a boundary layer forms for wider structures, lowering the elastic energy accumulation below the NR theory approximation. The exact boundary layer solution for the catenoidal case of $\phi_0=0$, eq. (\ref{eq: excesscat}), is plotted in \ref{fig:flattening}(a), showing better agreement with the discrete SWM numerics. \cite{tyukodiinprep2022}
%  {\bf cite Tyukodi et al, in prep.}
 Notwithstanding the discrepancy at intermediate scale, the flattening cost, $E_\infty= E_{\rm ex} (w\to \infty)$, is the same in the narrow-ribbon approximation and boundary layer solutions as indicated on the right of \ref{fig:flattening}(a). The difference in energies with $\phi_0$ for large $w$ is interpreted to be largely due to the differing values of $E_\infty$. Some numerical results at small $\theta_0$, in the case of $\phi_0 = 45^\circ$ were at unexpectedly large energy after minimization. This is attributed to the limited resolution of the minimization for the smallest values of $\theta_0$, which reach low force tolerance to meet the stopping criterion without fully resolving the small residual strains in the structure.
 
 The elastic energy due to shape flattening $E_\infty$ in the limit $w \to \infty$ was further analyzed by minimization of tubule assemblies of SWMs. Closed tubules were prepared with monomer bonds aligned to minimize the bending energy, e.g.  monomers with $\phi_0 = 0$ had bonds aligned parallel and perpendicular to the tube axis whereas for $\phi_0 = 45^\circ$ the bonding directions are at $45^\circ$ with respect to the tube axis.  For intermediate $\phi_0$ values we analyzed the higher pitch helical geometry, but confirmed that both branches are degenerate in the $w \to \infty$ limit (i.e. in the limit of vanishing boundary layer contributions). To account for the boundary layer relaxation of finite-length tubes, tubes of varying length were minimized, from $100 d$ to $150 d$ in increments of $10 d$.
 The flattening energy was then found by extrapolation to infinite length. The results for five values of $\phi_0$ and varying $\theta_0/r_a$ are shown in Fig.  \ref{fig:flattening}(b). The dependence of flattening cost on $\phi_0$ via the anisotropic bending costs is captured by the harmonic approximation, eq. (\ref{eq:gb_einfty}). However, results with significant strains associated with flattening, $\delta r \approx t \theta_0 / 2$, relative to the range of interaction $r_a$, show a reduction in the flattening cost due to strain softening as the interaction potential drops significantly below its harmonic approximation (see Fig. \ref{fig:monomer_design}(a)). The modified predictions for flattening cost, plotted for varying $\theta_0 d/ r_a$ as dashed curves in Fig.  \ref{fig:flattening}(b), are computed by minimizing attractive interactions over monomer orientations while enforcing uniform flattening with monomers maintaining spacing $d$ (i.e. numerical minimization of eq.~(\ref{eq: bendSWM}) over $C_{ij}$ subject to $K_G = 0$ using the fully non-linear form of soft attractive potential). The flattening geometry for $\phi_0 = 0$ and $ 45^\circ$ is illustrated in \ref{fig:flattening}(c), where the monomers in the tube are colored by excess energy to show the significant strain relaxation near the boundary and uniform strain in the interior. The comparison between (cylindrically) flattened and (hyperbolic) target geometry are illustrated for a $9\times 9$ cross array of SWMs blocks.  Notably, this highlights that shape flattening for catenoidal SWMs ($\phi_0=0$) membranes generates row unbending, whereas for helicoidal SWMs ($\phi_0=45^\circ$) rows are untwisted from their target binding.  The combined effects of shape flattening transitioning from unbending to untwisting as $\phi_0$ increases with a greater twist stiffness than row-bending stiffness ($B_{\perp} = 2 B_{\parallel}$), leads to the ($\sim 30\%$) increase in elastic shape-flattening energy from catenoidal to helicoidal assembly observed in Fig. \ref{fig:flattening}(b).
 
 To summarize, ring and ribbon morphologies of SMWs exhibit ground-state energetics that are well-described by the NR theory as summarized in Sec.~\ref{sec: continuum}.  Additionally, we find that the wide-ribbon regime, where shape-flattening leads to saturation of the frustration cost, is also well described to a first approximation by the continuum model, eq. (\ref{eq:gb_einfty}), although strain softening affects reduce this energy by up to $\sim 10\%$ for large wedge angles. Building from these results, we consider the zero-temperature thermodynamics of width limitation in the next section.

\subsection{Self-limitation and range of size control}

\begin{figure*}[h!]
% \centering\includegraphics[width=1.0\linewidth]{fig_sla_curves.jpg}
\centering\includegraphics[width=0.7\linewidth]{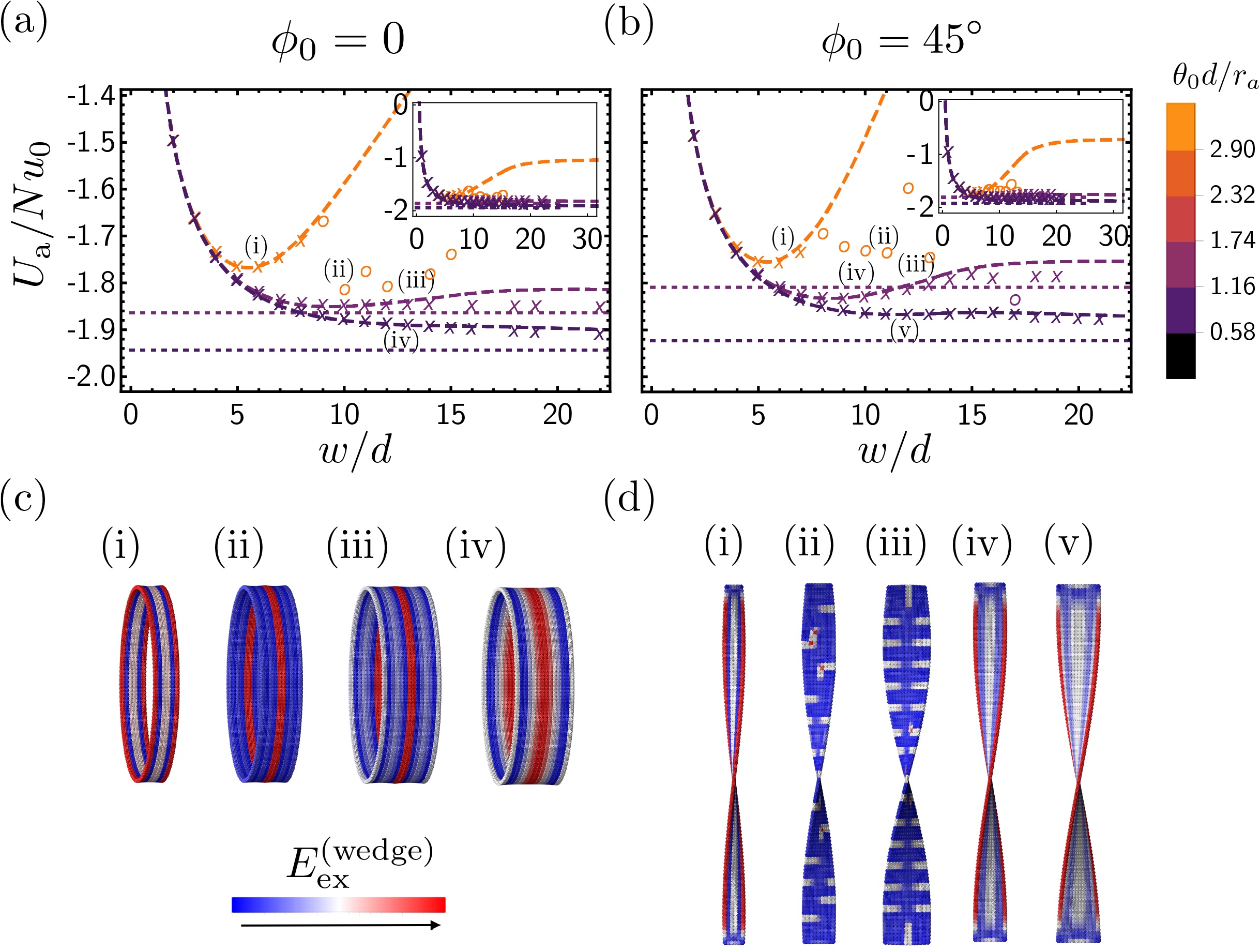}
\caption{ Self-limiting assembly and yielded states. (a-b) Average energy density $U_{\rm a}/N$ of minimized structures shown as points, for $\theta_0 = 2.5^\circ$ and $r_a/d = 0.014$, 0.036, and 0.057. Continuum theory predictions with harmonic elasticity are shown as dashed curves and nonlinear flattening energy predictions are shown as dotted curves only for the cases where flattened state is stable to yielding. Both size selection behavior and the transition to escape show agreement with theory, but yielded structures denoted with open markers in (a-b) exist that have similar or even lower energy. 
% \MJS{check change in this sentence}. 
(c-d) Typical structures are shown, with monomers colored by relative potential energy $E^{\rm (wedge)}_{\rm ex}$. Structures (ii) and (iii) are yielded for both (c) and (d).
} \label{fig:sla_curves}
\end{figure*}

By including the effect of edge energy due to missing bonds at the assembly boundaries (i.e. the effects of $E_{\rm edge}$), we develop predictions for the possible equilibrium self limitation in the SWM model in the limit of zero temperature. The competition between surface energy and super-extensive elastic energy may result in minima in the energy-density landscape $U(w)/A$ at finite $w$. This minima is the self-limiting state, when its energy is less than the bulk flattened state $U_\infty/A = E_\infty/A$, which in this case is a self-closing tubule.~\cite{hagan2021}
In Fig.  \ref{fig:sla_curves}, typical data from minimizations for varying $\phi_0, \theta_0, r_a$ are shown. The corresponding linear-elastic, narrow-ribbon continuum model predictions are shown as dashed curves, along with flat dotted lines for the prediction of $E_\infty$ according to the strain-softened flattening (bend) energy shown in Fig. \ref{fig:flattening}(b) . Typical SWM ground state structures are rendered in Fig.  \ref{fig:sla_curves}(d-f), for $\phi_0=0^\circ, 22.5^\circ$ and $45^\circ$.

%MJS moved to get on p 10

We start by noting the existence of a well-defined minimum at $w_0=(5\mbox{-}6)d$ for the smallest attraction range  ($r_a = 0.014 d)$, or equivalently higher stiffness to cohesion ratio $Y / \gamma \propto r_a^{-2}$, for both catenoidal and helicoidal assemblies, with quantitative agreement between discrete SWM model and the continuum model.  These minimal energy states fall well below the expected flattening energy from continuum theory, suggesting the thermodynamic ground state has frustrated-limited finite width~\footnote{$E_\infty$ is not shown in  Fig. \ref{fig:sla_curves}  for the cases with smallest $r_a$, when the corresponding strains exceed the inflection points of the attractive potential, $\epsilon \approx \theta_0 t /2 = r_a / 2$.}.  Notably, and as discussed in detail in Section \ref{sec: bond yielding} below, ground state structures for larger widths than the minimum fall off the curve predicted by the continuum theory, an effect which can be attributed to non-linear yielding of highly strained SWM bonds (locally yielded bonds appear as high-energy density bands in Fig. \ref{fig:sla_curves}(c-d)).

%$E_\infty$ is not represented for the cases with smallest $r_a$, when the corresponding strains exceed the inflection points of the attractive potential, $\epsilon \approx \theta_0 t /2 = r_a / 2$. When this criterion is met, the tube structure may be unstable to breakup into smaller structures. For the finite-width structures, the attractor separations between adjacent monomers were analyzed for the minimized structures, and structures where any attractor bond had separation exceeding $r_a / 2$ are plotted with open markers in the results. The yielded structures show distinct rows of high-energy in renderings in \ref{fig:sla_curves}(d-f). These yielded structures deviate noticeably from the continuum model predictions, and can even achieve lower energy density $U/A$ than the expected minima. The role of yielded structures on equilibrium self-limiting assembly is considered in the next section.

%Among the structures without yielding, the discrete model numerics show good agreement with the model predictions. At small $r_a$, or equivalently higher stiffness to cohesion ratio $Y / \gamma \propto r_a^{-2}$, an energy density minimum is present and the finite state is of substantially lower energy than the bulk state. This is illustrated in the insets in figure \ref{fig:sla_curves}, extending the model energetics predictions to a broader range of $w$, where the finite-$w$ minimum may be favored over the tubule state by energies on the order of $u_0/2$. 

For larger values of $r_a$, the SWM ground states show general agreement with the predictions of the continuum model, including an optimal $w_0$ and energy substantially increasing for successively larger $w$ notwithstanding the instability associated with yielding for large enough structures. With increasing $r_a$, which leads to effectively softer assemblies, the minimum becomes more shallow and eventually metastable to the defrustrated (i.e. $K_G \to 0$) state. Metastable minima can be resolved for increasingly soft assembly parameters, but cannot be resolved beyond $w = w_*$ despite minimizations extending up to $w = 2 w_*$. Self-limiting minima beyond $w_*$ are not predicted for the NR continuum model(see \ref{section:si_continuum}).

\begin{figure*}[h!]
\centering\includegraphics[width=0.65\linewidth]{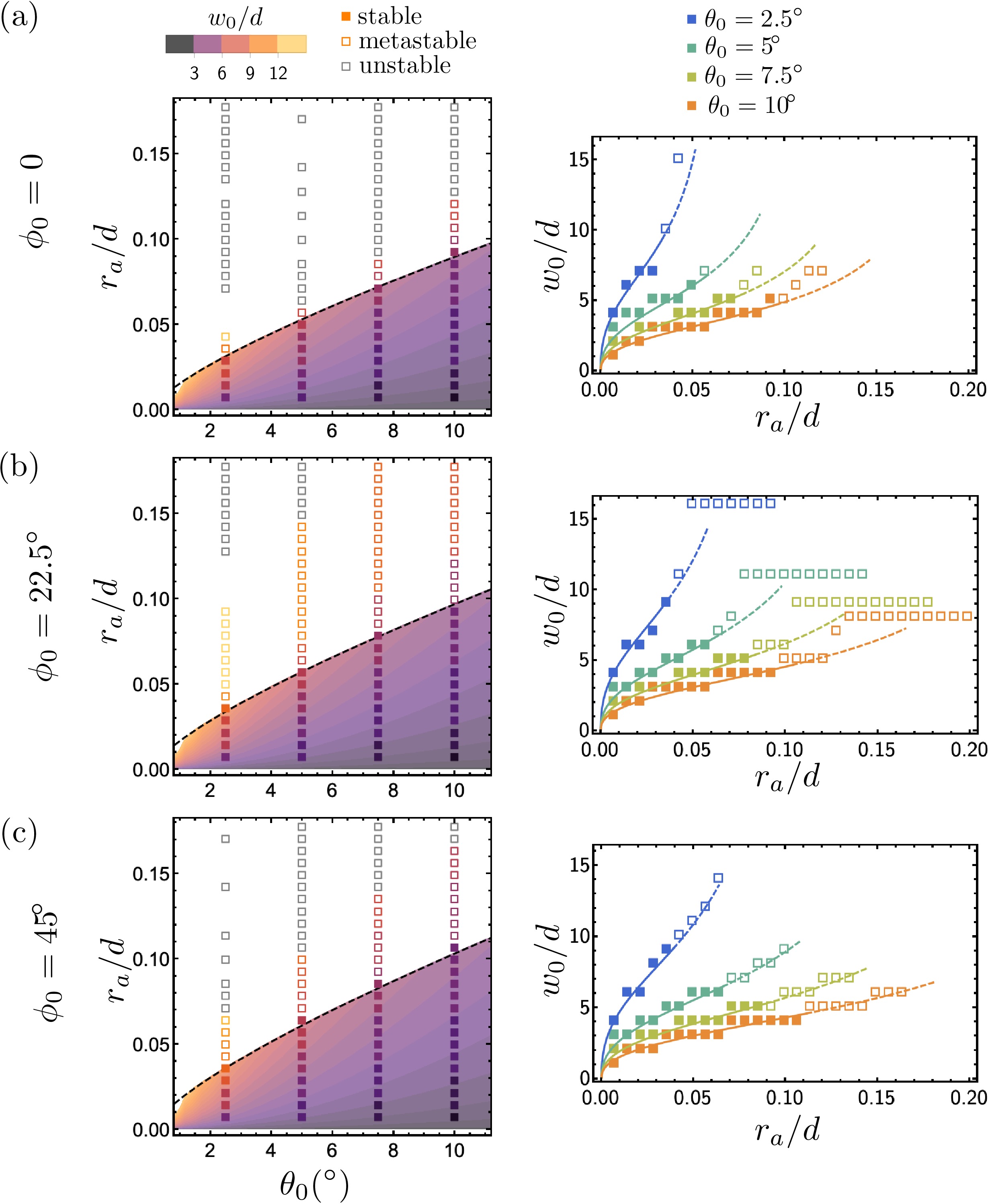}
\caption{ Phase diagrams and range of self-limiting assembly. (a-c) Left column shows self limitation phase diagrams in the design space of interaction range relative to monomer size $r_a/d$ and amount of monomer shape misfit, monomer wedge angle $\theta_0$, for three values of $\phi_0$, where points are colored according to the optimal assembly size $w_0/d$ or gray if there was no minimum up to $2w_*$. Open markers indicate that the flattening energy prediction $E_\infty$ was lower than energies of minimized finite-size assemblies. The continuum prediction is plotted with contours. On the right column, the same data is plotted to show predicted size $w_0/d$ as a function of interaction range $r_a/d$ for curves of constant misfit $\theta_0$. Curves represent prediction from continuum theory with solid curves where minima are stable and dashed for metastable states. 
% The linear theory prediction for $w_{\rm max}$ is denoted with 'X' \MJS{I don't see an X} and the predictions with nonlinear flattening energy are open markers.
} \label{fig:sla_phase}
\end{figure*}

The predictions for equilibrium self-limiting size are presented in Fig.  \ref{fig:sla_phase}, for the full range of parameters investigated. Specifically, we denote SWM assembly {\it self-limiting} for parameters where we resolve a local minimum in the energy density {\it and} that minimum falls below the predicted shape flattening energy density.  For this range of SWM taper angles ($\theta_0 \geq 2.5$), we find the range of size control agrees with the continuum model (including strain-softening corrections to $E_\infty$), with equilibrium sizes up to $9$ monomer widths in length, consistent with $w_0 \leq w_*$~\footnote{The boundary layer corrected prediction for moderately larger sizes achievable for $\phi_0 = 0$ is not apparent, and may be too small of an effect to appear in the discrete model}. Taken together, these show that anharmonic (i.e. strain-softening) effects of bonds in the discrete SWM model,  which having little effect on the small, finite-width assembly energetics, lead to measurable reductions in the range of size control, relative to purely linear-elastic model descriptions.

\subsection{Role of bond yielding in self-limiting assembly} \label{sec: bond yielding}

The structures with large internal strains shown in Fig.  \ref{fig:sla_curves} are a consequence of bond yielding, which occurs at the inflection point in Fig. \ref{fig:monomer_design}. In general, models of self-assembly with geometric frustration and also with finite-range interactions can exhibit regimes where the internal strains associated with stress accumulation are greater than the interactions can support. This can result in a distinct mode of thermodynamic escape from self limitation, such as the nucleation of low-symmetry, cracked assemblies in curvature frustrated tubules.\cite{tyukodi2021thermodynamic} In this regime, complex branched morphologies are expected, which are composed of stronger bound and elastically coupled regions, weakly bound together by partially yielded, yet at least slightly cohesive zones, as we observe in the partially yielded SWM ground states in Fig. \ref{fig:sla_curves}(c-d).

\begin{figure*}
% \centering\includegraphics[width=0.9\linewidth]{figure_rings_stacking.jpg}
\centering\includegraphics[width=0.9\linewidth]{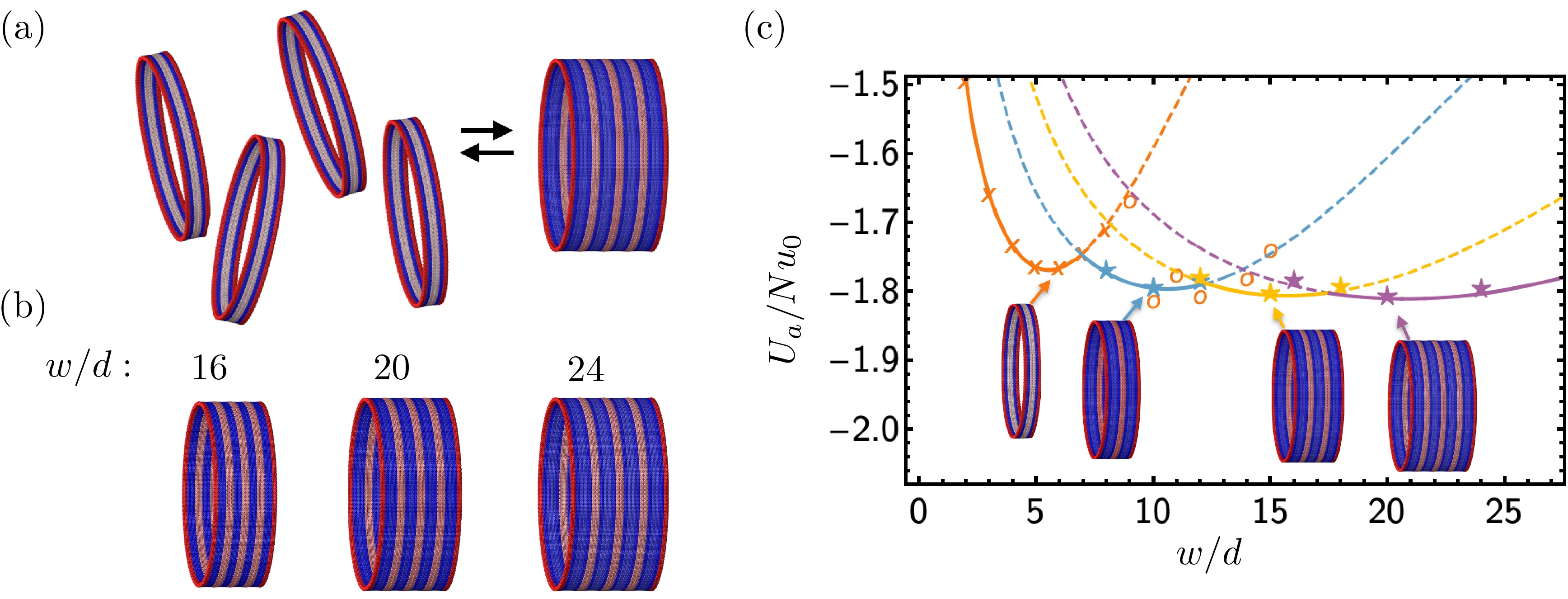}
% \centering\includegraphics[width=1.0\linewidth]{fig1.png}
\caption{ Role of highly deformed bonds in hierarchical assembly. (a) The self-limiting structures of optimal width, studied here, are expected to coexist with stacked structures, which gain some additional cohesion. (b) Pre-built structures of width $w$ = 16, 20, and 24 $d$ are generated from the previously relaxed structures of width $w$ = 4, 5, and $6 d$ at $r_a/d = 0.014$, $\phi_0 = 0$, and  $\theta_0 = 2.5^\circ$. (c) The energies of pre-yielded structures are plotted versus the width with curves generated from the non-yielded harmonic theoryincluding an additional contribution from yielded bonds fit to $u_\mathrm{yield} = 0.15 u_0$. } \label{fig:yielded_rings}
\end{figure*}

To further rationalize the energetics of yielded structures, we consider the simpler case of the low-energy yielded structure (ii) from Fig.  \ref{fig:sla_curves}(d), a catenoidal ring. Here, the yielded bonds are approximately in the middle of the structure, and the structure of total width $w/d = 10$ is nearly twice the width of the self-limiting structure at width $w/d = 6$. We consider this structure as a composite of two non-yielded structures with yielded bonds acting as weak, partial bonding between the two structures. For this particular monomer geometry and parameters, we further explore the energetics of partial-bonding between self-limiting rings, with results presented in Fig.  \ref{fig:yielded_rings}. Additional minimizations were conducted, starting with the minimized and non-yielded structures of width $w/d = 4$, 5, and 6. For each starting width new structures were prepared by arranging stacked copies of that structure, such that the copies did not interpenetrate but partial bonds were made, where one of the four attractors on a binding face coincided with its neighbor. After further minimization to the same force tolerance $0.3 \times 10^{-4} u_0/d$, the results are presented, with colors corresponding to the number of copies. The energetics of this composite ring morpholgies are well-described by an augmented model that uses equation \ref{eq:model_energy} to describe the energies of the constituent non-yielded rings and fits a constant energy of $0.15 u_0$ to each of the partial bonds. These partially bonded structures are of lower net cohesive energy than the isolated, non-yielded structures. 

More general considerations (i.e. at least weakly cohesive binding between elastically coherent self-limiting membranes) imply that such hierarchical morphologies are possible for any SWM assembly, at least at sizes sufficiently larger than $w_0$, and generically such structures should have at least slightly lower total energy than the elastically self-limiting states (i.e. associated with the minimum in the energy density).  However, due to the relatively weaker cohesive energy binding the structures at these yield bonds, it is expected these hierarchical structures may be broken up due to entropic considerations at sufficient high temperature, leading to an equilibrium state dominated by the elastically self-limiting morphologies identified in the phase diagram shown in Fig.~\ref{fig:sla_phase}.

\section{Discussion and conclusions}

In summary we have developed and studied a discrete SWM model of hyperbolic membrane assemblies with crystalline order.  Detailed analysis of the minimal energy density morphologies is compared to predictions of linear-elastic, continuum theory illuminating the connection between microscopic features of the frustrated particles and their mesoscopic structure and thermodynamics.  To conclude we discuss the implications of these results for the understanding and engineering of self-limiting assembly of geometrically frustrated building blocks.

\subsection{Controlling self-limiting dimensions through discrete building block shape and interactions}

In this work, we identify the role of extrinsic geometry to shape equilibrium self-limitation via, $\phi_0$, the angle relating bonding directions to the direction of rotation between bonded monomers. While the SWM  model realizes equivalent stress accumulation, $E \sim Y \kappa_0^4 w^4$ independent of $\phi_0$, the elastic cost of escaping frustration through shape flattening can vary with $\phi_0$ when there is anisotropic bending cost (i.e. $B_\perp \geq B_\parallel$)~\footnote{Notably, while the priori continuum analyses ~\cite{Armon2014} are based on isotropic bending elasticity ($B_\perp = B_\parallel$) the anisotropic case $B_\perp \geq B_\parallel$ is fairly generic for purely attractive binding geometries, as twist deformations load all cohesive bonds in proportion to their distance form the rotation axis, whereas bend distortions only load bonds in proportion to projected distance to the neutral axis of bending.}. For this reason, the extrinsic geometry has the potential to influence assembly energetics because $\phi_0$ relates the directions of preferred curvature to the bonding directions, determining the extent to which different modes of bending are activated with associated moduli $B_\perp, B_\parallel$.  This is most clear in dependence of flattening energy $E_\infty$ on $\phi_0$ shown in Fig.~\ref{fig:flattening}b, which derives from the transition from unbending ($\phi_0 =0$) to untwisting ($\phi_0 = 45^\circ$) in the flatten tubule geometry.  A consequence of the increasing escape energy with $\phi_0$ is that the flattening size $w_*$, and to a smaller extent $w_{\rm max}$, are predicted by NR theory to {\it decrease} with $\phi_0$, as shown in Fig.~\ref{fig:contin_wmax}c.  A countervaling trend might be anticipated, based on the previously reported effects of the boundary-layer corrections on the shape-flattening tranisition~\cite{Armon2014} for equal bend and twist constants ($B_\perp=B_\parallel$), as shown  
% \MJS{ there is no 2d}\DMH{fixed}
Fig.~\ref{fig:contin_wmax}d, which shows that $w_*$ instead {\it increases} by $\sim 20\%$ with $\phi_0$.   To assess whether the shape-flattening width should decrease with increasing $\phi_0$ as suggested by NR theory or whether boundary layer corrections dominate and lead to the opposite dependence, we extracted, we extracted the apparent $w_*$ (for $\phi_0 \neq 0$) and $w_{\rm max}$ from SWM ground state simulations.  Plotting these in Fig.~\ref{fig:contin_wmax}c, we note that there may be a slight tendency for $w_*$ to decrease with $\phi_0$, but this is obscured by the resolution limits imposed by discreteness of the changes in $w$ possible for the SWM model.  We observe no measurable changes in the maximum self-limiting size with curvature angle.  Hence, notwithstanding these two possible mechanisms for $\phi_0$ dependence, the size-range of self-limitation and stress-accumulation appear to relatively insenstive to intrinsic geometry of the SWM ribbon morphology.

% Beyond the role of extrinsic geometry (i.e. the direction of curvature axes), the SWM model highlights special dependence of self-limitation on the brittle nature of inter-subunit forces. \MJS{Why brittle? Is brittle intrinsic?} 
%  - we won't use brittle since this paragraph is about stiffness not modes of material failure
Beyond the role of extrinsic geometry (i.e. the direction of curvature axes), the SWM model highlights the special dependence of self-limitation on the range of interactions via the stiffness associated with accumulating frustration costs to the assembly.  
% In effect, the attractive range $r_a$ compared to SWM size $d$ controls whether assemlbies are effectively brittle (small $r_a$) or ductile (large $r_a$), which in turn, dictates the effects of geometric frustration in the assembly.
Consistent with the generic predictions of continuum theory, results in Fig. \ref{fig:sla_phase} confirm that the {\it maximum} self-limiting width, $w_{\rm max}$, for a given block geometry is dictated by the flattening size scale, $w_{\rm max} \lesssim w_* \propto (B/Y)^{1/4} \kappa_0^{-1/2}$, simply because the energetics approach extensive scaling in this limit.  In the simplest case, where interactions are purely cohesive, the ratio of bend to stretch ratio is controlled by the thickness (i.e. $B/Y \propto t^2$) which itself is of order of the particle size $d$.  Hence, this suggests the maximum self-limiting size $w_{\rm max} \approx d /\sqrt{\theta_0}$, which implies that the self-limiting dimensions that far exceed the size of the building block require small taper angles, $\theta_0 \ll 1$.  Additionally, the relations determining equilibrium size $Y \kappa_0^4 w_0^5 \approx  \gamma / w_0$ \ imply that that decreasing the degree of frustration through the wedge angle also decreases the range of edge energies for equilibrium, self-limited structures, which are characterized by the {\it maximum} edge energy $\gamma_{\rm max}/Y \approx  \kappa_0^4 w_{\rm max}^5 \approx d \theta_0^{3/2}$.  Notably, the ratio $\gamma/Y \propto r_a^2/d$ is a {\it cohesive elastic} length scale, which is most strongly dependent on the range (more strictly, the stiffness) of the cohesive interactions between subunits.  Taken together, these two relations show that decreasing the degree of frustration through reduced wedge angle increases the size range of frustrated limited assembly, but does so at the expense of requiring narrowed range of interaction stiffness.  In particular, thermodynamic self-limitation by frustration is only possible for $r_a/d \lesssim (w_{\rm max} /d)^{-3/2} \propto \theta_0^{-3/4}$, implying that self-selection on larger, multi-subunit dimensions requires increasingly shorter range (i.e. stiffer) cohesive interactions, which is consistent with the shift to smaller $r_a$ range with decreasing $\theta_0$ shown in Fig. \ref{fig:sla_phase}.

Notably, this basic result is predicated on two assumptions: (i) that deformations in assembly primarily strain interactions while subunits are considered rigid and (ii) binding interactions are purely attractive.  Relative to the case considered here, including additional deformability in the subunits themselves as in models of Refs. \citenum{Lenz2017, tyukodi2021thermodynamic, meiri2021cumulative}, the expectation may be to reduce the cohesive-elastic ratio $\gamma/Y$ below the (upper bound) limited by interaction stiffness, such that for any {\it finite} interaction range, inter-subunit deformability should only further depress the feasible size-range.  On the other hand, it can be shown that more complex binding geometries, incorporating distinct spatial patterns of local attraction and repulsion can give rise to effective elasticity where $B/Y \gtrsim t^2$,~\cite{Spivack_2022} thereby extending the elastic scale where shape flattening takes place.

Applying these elementary considerations to the experimental $C_{12}$-$\beta_{12}$ gemini amphiphile system studied in Ref. \citenum{Zhang2019} where width-dependent helicoidal ribbons morphologies were carefully characterized, we note, of course, that macromolecular subunits are both highly deformable and realize highly complex interactions.  Nevertheless, we may assess the apparent effects of the likely ranges of interactions on the overall morphologies and likely equilibrium states.  In these experiments, at early times, helicoidal ribbons are observed with mesoscopic pitches of order $\sim 100$ nm.  Based on a considerations of local packing in the ``twisted crystal'' of amphiphiles, ~\cite{Zhang2019} It was estimated that $d \approx 0.6$ nm, $\kappa_0 \approx 0.03~ {\rm nm}^{-1}$.  Assuming the naive estimate $\sqrt{B/Y} \approx d$, this suggests a shape-flattening size scale $w_* \approx 40$ nm, and at this shape-flattening size, ribbons are of order $w_*/d \approx 60-70$ sub-units across.  Notably the estimate of $w_*$ is consistent with the fact that at longer times, as ribbons grow larger than this size range, they exhibit a shape-transition to spiral ribbons, consistent with the basic predictions of the continuum elastic theory, but ultimately suggesting that thermodynamic equilibrium does not correspond to a regime where frustration stabilizes the open-boundary helicoidal ribbon, presumably because the assembly is too ductile.  This raises a basic question: presuming cohesive interactions govern the elasticity of assembly, how close might such a molecular system be to an equilibrium state of frustration limitation?  Applying the estimate of the upper limit on interaction range for the gemini amphiphile membrane, we find $r_a/d \lesssim (w_*/d)^{-3/2}~10^{-2}$.  Notably, as subunits are molecular in dimension, this corresponds to a limiting interaction range that is sub-Angstrom, clearly much shorter range than what might reasonably be expected from van der Waals or hydrophobic interactions which bond the amphipillic subunits together.  That interactions likely far exceed this range is consistent with the observed long-time growth of chiral amphiphiles assemlbies into shape-flattened tubules at long times, and more generally suggests that the possibility of frustration-limited assembly in molecular crystalline membranes may be difficult to achieve, if at all possible. 

%One question is whether the system could be modified to stabilize self-limiting ribbons in equilibrium. Here, van der Waals interactions between head groups of the amphiphiles were connected to the intrinsic curvature.  It was estimated that $d \approx 0.6$ nm, $\kappa_0 \approx 0.03$rad/nm and $w_* \approx 40$ nm. From the continuum model, setting aside the moderate intrinsic mean curvature contribution found in this system and taking $B_\parallel=B_\perp$, one finds that $w_{\rm max} \approx 0.6 w_*$. The corresponding range of interaction needed to achieve self-limitation using the double-wedge model would be $r_{\rm max}/d \approx \sqrt{w_{\rm max}^5 \kappa_0^4 / 1440 d} \approx 0.09$. More generically, the interaction range parameter will characterize assembly stiffness $\|u''(r)/u(r)\| |_{r=r_0} = \pi^2 / r_a^2$, where $r_0$ is the value of $r$ minimizing $u(r)$. 

The restrictions placed on the elasticity and range of frustration accumulation by the interaction range suggest the feasible avenue for engineering self-limiting systems requires the combination of (larger) colloidal-scale particles bound by shorter range interactions. We point out a recent example of DNA origami particles \citenum{gerling2015dynamic, sigl2021programmable, hayakawa2022geometrically} designed to be triangular subunits, $\sim 50$ nm in size with controllable inter-particle geometry, and large $w \gg d$, assemblies driven by short-range base-stacking interactions, whose interactions range may be less than $\sim 1$ nm. \cite{kilchherr2016single} Beyond the requirement of shorter-range interactions relative to subunit size, frustration-limitation will also require engineered  colloidal particles with the combination of precise geometry binding and also stiffness comparable to or exceeding that from the short-range stacking interactions.

\subsection{Role of finite interaction ranges: hierarchical aggregation }

The results presented here show features of discrete systems that are not captured by the linear-elastic continuum description. Two possible features have been proposed to augment the continuum description. Firstly, we identified the relevance of nonlinear elasticity, which becomes relevant as the scale of deformations in the equilibrium structures $ \sim d \kappa_0 t$ becomes comparable with the range of interactions $\sim r_a$. Strain softening was shown to reduce the cost of tubule formation, with results presented in Fig.  \ref{fig:flattening} that were well-described by the nonlinear elastic description of bending costs. Secondly, as the scale of deformations approaches the range of interactions, structures become mechanically unstable when individual interactions reach the yield point of the interactions. For our soft-binding model this corresponds to $r_y=r_a/2 \approx d \kappa_0 t$. 
%MJS split sentence as too long
The zero-temperature optimal size $w_0$ presented in Fig. \ref{fig:sla_phase} do not show a deviation from theory due to the breakup of bonds at high strains. However, the weakly-cohesive aggregation of otherwise elastically deformed, cohesive and finite assembly domains suggested here (as well as the low-symmetry internal cracking exhibited in the model of Ref. \citenum{tyukodi2021thermodynamic}) constitute an alternative manner of {\it escaping} the self-limiting thermodynamic consequences of frustration, that will occur in any realistic, particle-based description of frustrated assembly. 

% In short, \MJS{what is this?} this follows simply from the simple argument, 
In summary, if there is a minimal energy density at width $w_0$, then two of these structures can at least weakly bind together (i.e. through partial yielding of cohesive bonds, or some set of bonds that do not transmit effects of frustration between domains) without introducing additional elastic costs into those two structures.  Hence, it is straightforward to argue that at $T=0$, the energy density of multiple %a $m$ 
weakly-aggregated domains of size $w_0$ will fall at least slightly below the single domain minimum.  This simple argument suggests that generically even in states where a self-limiting domain minimum falls below the energy of the (smooth) shaped-flattened states, at sufficiently low temperature (and high concentration), self-limiting aggregates would be unstable to some condensed, multi-aggregate morphology that is effective unlimited in size, and points to the importance of finite-$T$ entropic effects in stabilize any putative regime of self-limitation. 
%\MJS{What do the MD simulations show? Does the internal boundary stress decrease, while the external boundary is the largest and the location of instability?} - \DMH{I think we can't say too much about the internal boundary stress- There's a lot happening in the disassembly and the stresses in the assembled cluster only "make sense" after averaging over time.}

% \begin{figure}[h!]
% \centering\includegraphics[width=1.0\linewidth]{fig_md_disassembly.png}
% \caption{ Dynamical disassembly/melting, evidence of hierarchical assembly transitions. Here is a placeholder, would like to measure melting temperatures (more likely, sequence of runs at very temperature to see short-time stability of stack) associated with stacked rings to isolated rings, then to monomers.} 
% \label{fig:md_disassembly}
% \end{figure}

\subsection{Modeling frustrated assemblies at finite temperature }

While the core analysis of this study focuses on purely energetic ground states at $T=0$, the conclusions detailed above imply some important open challenges for understanding frustrated self-assembly at finite temperature.  

For one, while the frustrated assembly seemingly offers an attractive paradigm for controlling equilibrium self-limiting assembly through engineering misfit of building blocks, ~\cite{Grason2017} we find that realizing such self-limitation at non-trivial size scales places important restrictions on the range of interactions.  In particular, thermodynamic conditions where frustration-limited morphologies out compete shape-flattened (i.e. defrustrate) morphologies rely on attractive interactions between subunits that are short-ranged, ideally much shorter-range than the size scale of the SWM subunits. 
%MJS but this does correspond to molecular systems such as DNA origami as noted above!
Using molecular dynamics to sample equilibration of discrete particle models, like SWM, in the self-limiting regimes faces additional challenges. Not only are the time scales to equilibrate such systems long, but stiff potentials require shorter time steps increasing the total computational cost. In addition, simulations of assembly of free particles will have a short capture radius, which will increase the simulation time.~\cite{Cheng2012, Hagan2014}

\begin{figure}[h!]
% \centering\includegraphics[width=1.0\linewidth]{fig_md_assembly.png}
\centering\includegraphics[width=1.0\linewidth]{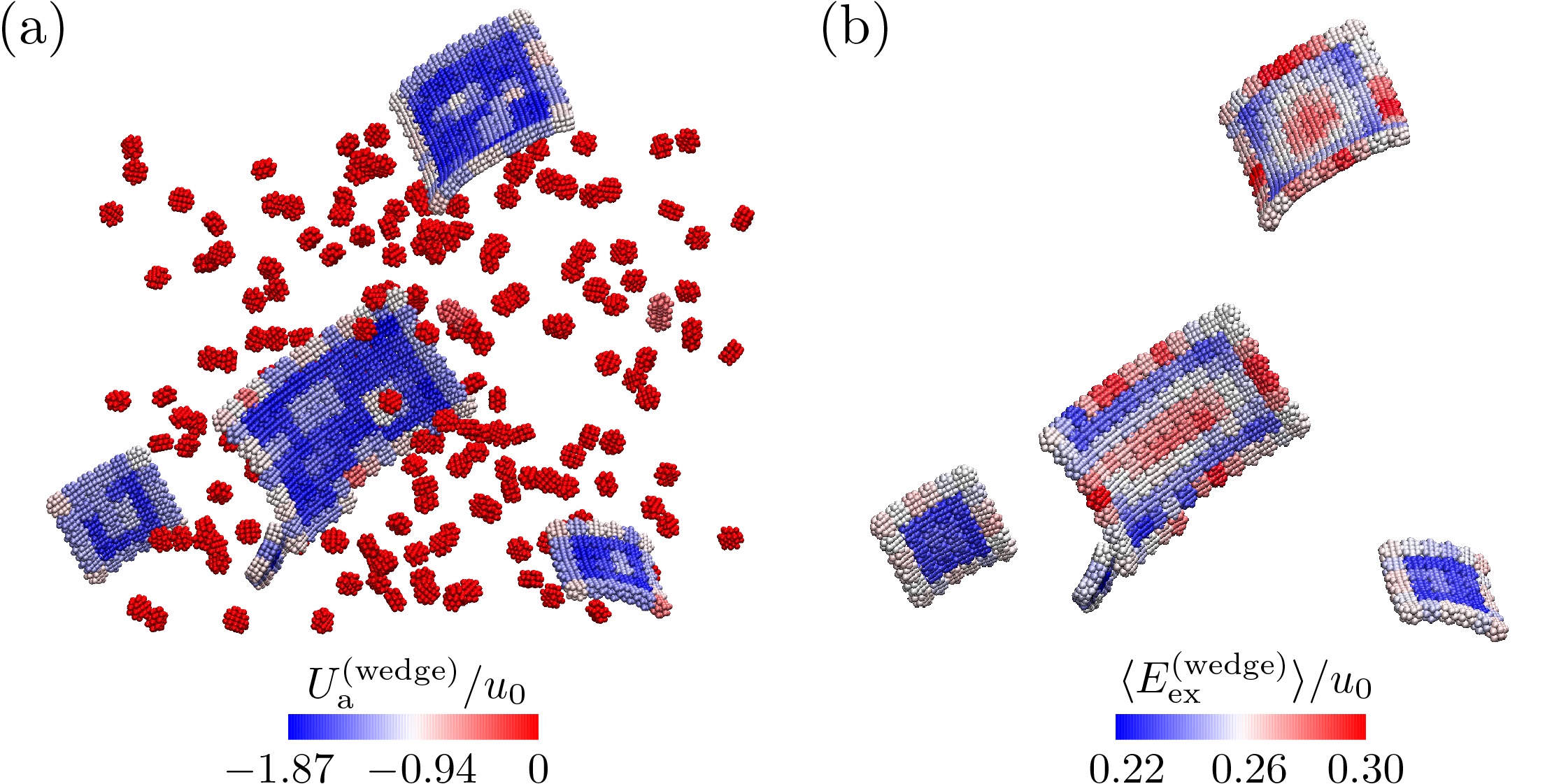}
\caption{ (a) Monomers with angles $\phi_0 = 0$ and $\theta_0 = 10^\circ$ assemble into clusters at moderate interaction range $r_a/d=0.14$ and temperature $k_{\rm B} T/u_0 = 0.054$. The clusters have edges predominantly oriented in the closed-packed directions. (b) Averaging the interaction energies over many timesteps reveals the gradients in strain energy that develop in clusters.
% \MJS{No (c). cut text.}
%(c) MD runs of the stacked-ring structure with yielded bonds, maintaining the short interaction range $r_a/d = 0.014$, show that the structure holds together except at much higher temperatures, $k_{\rm B} T/u_0 = 0.064$.
%(a) At lower frustration, see anisotropic patches. Note that clusters have edges aligned with close-packed directions, i.e. surface energy is anisotropic, which would be a necessary feature for different behavior with different $\phi_0$. At $T=0$ these parameters would favor escape to tubule. (b) At higher frustration, cluster is highly branched with branch width $~4d$. At zero temperature, expect $w_0=3d$. (c) Reducing temperature, find self-closing clusters, close to zero-temperature preferred ring circumference of $18 d$ but still have branching. The zero-temperature optimal size is still $w_0=3d$.
%\{second option for this figure}
%Lower row: For the system $\phi_0 = 0, \theta_0 = 20^\circ$ (a) Zero temperature energy landscape shows optimal size $w_0 = 3d$ for $r_a/d = $ (b)
% Top: MD assembly Bottom: stacked ring is stable at T = 0.8$\epsilon$ but begins to disassemble at T = 0.9$\epsilon$.
} 
\label{fig:md}
\end{figure}

% \MJS{these simulations are for a large $r_a$. Ideally I would like to have a shorter value to say that we can at least see assembly. I will try to run them.}

As a preliminary picture into the finite temperature self-assembly behavior of the discrete particle model, we carried out MD simulations of SWM model starting from disassociated configurations. In these simulations, 500 free subunits are randomly placed in the simulation box and the temperature is $0.8 \epsilon$. The simulation system has 500 subunits at a density of $6.35 \cdot 10^{-4} \, {\rm particles}/\sigma^{3}$, and the subunit parameters are $\phi_0=0, r_a=0.5\sigma$ (i.e. relatively large range) and $ \theta_0=10^\circ$. Fig. \ref{fig:md} shows that intial assembly does occur for this system in this time window, with wultiple rectangular assemblies have formed after 1 billion time steps. While further work is needed to determine the equilibrium structures and more fully explore the kinetics assembly, it clear that SWM parameters can be identified where MD simulation is viable.

 % more about the MD assembly part
 Notably, while the free SWM simulations are free to sample much more irregular morphologies, the assembled clusters shown in Fig. \ref{fig:md}(a) are mostly rectangular with boundaries aligned along the close-packed directions, which is favorable for cohesive assemblies to optimize the number of bonds formed for a given number of particles. This feature suggests the possibility that larger assemblies will maintain boundaries along the close-packed directions, which was implicit in the choice of pre-assembled structures of varying-$w$ that were generated for the $T=0$ results presented here. Coloring monomers by their respective interaction energy in Fig. \ref{fig:md}(a) reveals that in addition to bonds forming and breaking, there are varying degrees of strain in the interior of each cluster, evidenced by patches of lighter color corresponding to more strained bonds. Further analysis revealed gradients in strain within the clusters assembled at finite $T$, shown in Fig. \ref{fig:md}(b). These more regular patterns of coherent strain gradients were extracted by averaging local binding in clusters over multiple time steps. Time sampling was chosen to be sufficiently short with respect to the lifetime of bonds but longer than the apparent correlation of elastic fluctutations or phonons within the structure.  The appearance of these coherent strain gradients, which grow in magnitude with size, are consistent with the fact accumulation of frustration costs in the SWM model shape finite temperature pathways.
 
 What remains to be determine are how these effects of strain accumulation also influence the kinetics of reaching equilibrium states, and further, whether and how finite temperature effects shift the expects phase boundaries between equilibrium self-limiting states and states of bulk assembly.  Not only do we expect that especially soft and weakly frustrated systems to escape frustrated to unlimited, shape-flatted structures, but we also expect, based on the analysis described in Sec. \ref{sec: bond yielding} that a low temperatures, otherwise frustration limited structures may condense or aggregate into (presumably low-symmetry) clusters held together by partially yeilding bonds.  Hence, a further open challenge is to identify how non-linear features of the inter-particle binding that control yield, shape the critical temperature and concentration conditions at which translation entropy favors break-up (or melting) of heirarchical clusters into free, size-controlled aggregates.

%We also briefly consider the effect of partial bonds to finite-temperature behavior in the double-wedge model. As shown in figure \ref{fig:md}(c), we further consider the stacked-ring structures with energetics characterized in figure \ref{fig:yielded_rings}, with weak bonds estimated to be $\approx 0.15 u_0$. The same yielded structure is run with MD at varying temperatures to assess stability over simulation times of 30 million time steps. We find that the stacked rings are stable up to $T=0.8\epsilon$. The local energy is very different from the minimized state as shown in figure \ref{fig:md}. In particular the variation of energy is smaller than in the minimized structure suggesting that the interring binding is less stressed. 
%At $T=0.9\epsilon$ the system begins to disassemble in the time frame simulated. The disassembly occurs only from the outer boundary with subunits diffusing away. We did not observe the multiring structure splitting into separate individual rings.

%Expect $\sim 3 k_{\rm B} T$ translational entropy to release a single ring, but weak binding energy is $0.15 u_0 \times (2 \pi d / \theta_0)$ i.e. scales with structure size via the circumference (??). Melting to monomers with $\sim 3 k_{\rm B} T$ versus $2 u_0$ per monomer (??). We conclude on this framework for control hierarchical assembly, where future monomer designs may favor weak bonding to a greater or lesser extent.

\section*{Acknowledgements}
The authors are grateful to M. Wang, B. Tyukodi and N. Hackney for valuable discussions on this work.  DH and GG acknowledge support for this work through US National Science Foundation through award NSF DMR-2028885, as well as through the Brandeis MRSEC on Bioinspired Materials NSF DMR-2011846. 
This work was performed, in part, at the Center for Integrated Nanotechnologies, an Office of Science User Facility operated for the U.S. Department of Energy (DOE) Office of Science. Sandia National Laboratories is a multimission laboratory managed and operated by National Technology \& Engineering Solutions of Sandia, LLC, a wholly owned subsidiary of Honeywell International, Inc., for the U.S. DOE’s National Nuclear Security Administration under contract DE-NA-0003525. The views expressed in the article do not necessarily represent the views of the U.S. DOE or the United States Government.
Simulation studies of SWM model were performed on the UMass Cluster at the  Massachusetts Green High Performance Computing Center and computers at Sandia.

% The Acknowledgements come at the end of an article after Conflicts of interest and before the Notes and references.

%%%END OF MAIN TEXT%%%


%%%REFERENCES%%%

\bibliography{wedge_rfs} %You need to replace "rsc" on this line with the name of your .bib file

\appendix

\section{SWM design details and pairwise elasticity} \label{section:si_monomer_design}

This section details the SWM geometry, the interactions defined by repulsive and attractive potentials, and the resulting elastic response of a bound pair of monomers. The SWM geometry is designed so that pairs of bonded monomers prefer to adopt a target configuration with opposite rotation sense of neighbors in the two bonding directions, a discrete analogue to local curvatures of a minimal surface and the rotation of that surface's normal vector. Here, the monomer's coordinate frame is denoted by unit vectors ${\bf c_1}$ and ${\bf c_2}$ that are orthogonal and point in directions of bonding faces 
(see the coordinate frame illustrated in  Fig.  \ref{fig:monomer_design} in the main text) and ${\bf c_3}$ is assumed to point along the local normal to the mid-surface spanned by the 2D assembly. 

\begin{figure*}[]
\centering\includegraphics[width=1.0\linewidth]{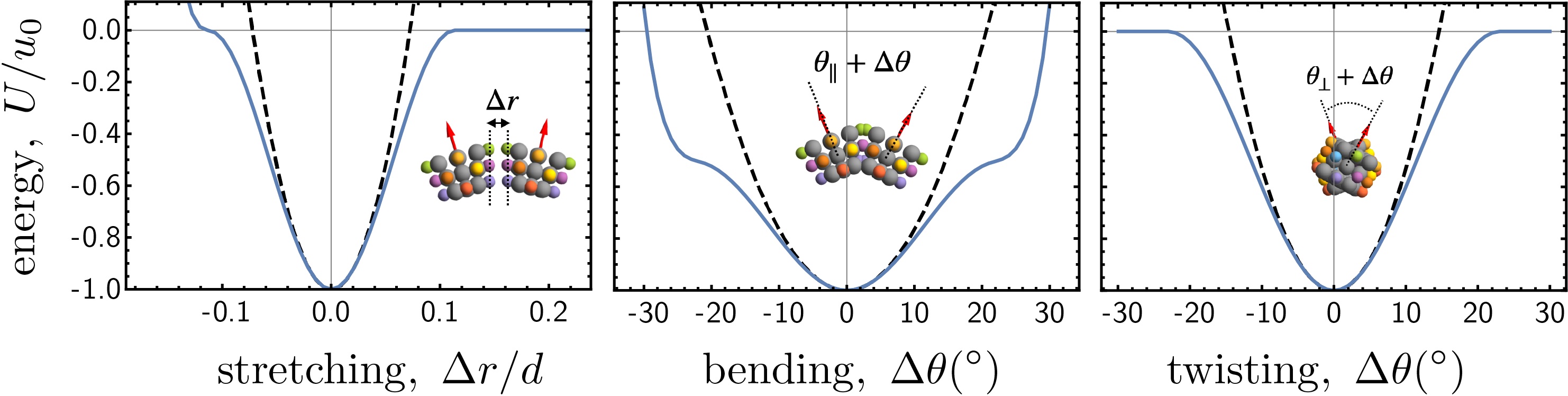}
\caption{ Pairwise elasticity due to attractor interactions. The energy of a pair of monomers for different deformations away from the preferred configuration is shown for $\theta_0 = 10^\circ, r_a = 0.11d$, with the harmonic approximation shown as a dashed curve. The repulsive interactions become nonzero at deformations much larger than seen for assemblies in this study, e.g. for compression $\delta r < -0.1 d$ or bending deformation angle $|\Delta \theta| > 20^\circ$. 
}
\label{fig:fig1a}
\end{figure*}

The rotation axes of the target configuration, relating a monomer's coordinate frame to those of neighbors bonded in the ${\bf c_1}$ and ${\bf c_2}$ directions are, respectively,
\begin{eqnarray}
    {\bf r_1}  \equiv - \sin 2 \phi_0 ~ {\bf c_1} + \cos 2\phi_0 ~ {\bf c_2},  \nonumber \\
    {\bf r_2}  \equiv \cos 2\phi_0 ~ {\bf c_1} + \sin 2 \phi_0 ~ {\bf c_2}.
\end{eqnarray}
In the target configuration, the frame of a monomer is related to its four neighbors' frames by a displacement along the bond direction $\frac{1}{2} d {\bf c_{\alpha}}$ ($\alpha = 1$ or $2$), rotation of angle $\theta_0$ about ${\bf r_{\alpha}}$, and additional displacement of $\frac{1}{2} d {\bf c_{\alpha}}'$ where ${\bf c_{\alpha}}'$ is the neighbor's frame after the rotation about ${\bf r_{\alpha}}$. Here, $d$ is the nominal monomer width related to the actual monomer geometry and the arrangement its of attractive sites, detailed below. To leading order in $\theta_0$, this transformation is a bending angle between neighboring monomers of $\theta_{\parallel} \equiv \theta_0 \cos 2 \phi_0$ projected along the bond direction ${\bf c_\alpha}$ and a twist angle $\theta_{\perp} \equiv \theta_0 \sin 2 \phi_0$ of rotation about ${\bf c_\alpha}$. The definition of ${\bf r_{\alpha}}$ enforces opposite sense of bending angle in the two bond directions, analogous to a minimal surface having curvature tensor with zero trace, and opposite sense of twist, which is true for a smooth surface that necessarily has symmetric curvature tensor. 

%The target configuration of monomers bonded in the ${\bf c_2}$ direction has an opposite sense, with displacement $d {\bf c_2}$ followed by rotation of $\theta_0/2$ towards each other about the vector $\cos 2\phi_0 {\bf c_1} + \sin 2 \phi_0 {\bf c_2}$, and correspondingly bend and twist angles are equal and opposite to those along the ${\bf c_1}$ direction.

The monomer design consists of an array of repulsive sites that define an excluded volume to monomer overlap, and attractive sites on four sides that determine the geometry of bound neighbors. This design is inspired by previous work on tubule-forming monomers in Ref. \citenum{cheng2012self, cheng2014self, stevens2017long}. Repulsive sites are of a single type and interact with the repulsive sites on other monomers according to the Weeks-Chandler-Anderson potential, 
%MJS fractions are small in array environment, need to put displaystyle for each row that has fractions. I also switched \Big to \left and \right but that does change anything
\begin{equation}
u_{\rm WCA}(r)= \left\{
\begin{array}{ll}
   4 \epsilon  \left[\left(\frac{\sigma}{r} \right)^{12} - \left(\frac{\sigma}{r} \right)^6 - \left(\frac{\sigma}{r_c} \right)^{12} + \left(\frac{\sigma}{r_c} \right)^6 \right], &   r \leq r_c\\
  0, &   r > r_c\\
\end{array} \right. 
\end{equation}
where $r$ is the distance between interacting sites, $\sigma$ and $\epsilon$ are the reference distance and energy units chosen for simulations. For this study, the soft attractive interaction was chosen so that $u_0 = 14 \epsilon $, consistent with attractive interaction strengths which favored finite temperature assembly in the tube-forming system. \cite{cheng2012self}
%MJS reordered sentences to discuss pair potential first, then discuss monomer-monomer interaction
The cutoff length $r_c = 2^{1/6} \sigma$ yields a purely repulsive interaction that reaches zero energy and zero force at $r = r_c$.
The nominal monomer width is related to $\sigma$ by $d = (2.4 + 2^{1/6}) \sigma$, which follows from the attractive site arrangement specified below. 

The arrangement of a monomer's 27 repulsive sites can be specified in terms of the orthonormal coordinate frame for a given monomer with ${\bf c}_1, {\bf c}_2$ the bonding directions and ${\bf c}_3 = {\bf c}_1 \times {\bf c}_2$. The arrangement varies with curvature angles $\theta_0, \phi_0$ to maintain separation between repulsive sites of bound neighbors that are close to the target configuration. Adopting indices $i_1, i_2, i_3$ that each take values $-1,0,1$, the repulsive site coordinates are 
\begin{widetext}

\begin{multline}
{\bf R}_{i_1, i_2, i_3}^{\rm rep} = \sigma \Big[%\langle
i_1 + i_1 i_3 \cos (2 \phi_0) \tan \left( \frac{\theta_0}{2} \right) + i_2 i_3 \sin(2 \phi_0)\tan \left(\frac{\theta_0}{2} \right) \Big] {\bf c}_1 \\ + \sigma \Big[i_2 - i_2 i_3 \cos(2\phi_0)\tan \left( \frac{\theta_0}{2} \right)+ i_1 i_3 \sin(2\phi_0) \tan \left( \frac{\theta_0}{2} \right) \Big] {\bf c}_2 +  \sigma \,i_3  \, {\bf c}_3%\rangle 
,
\label{eq:repcoords}
\end{multline}
\end{widetext}

Reference: https://www.physicsforums.com/threads/wide-equation-in-revtex-4-2-is-overlapping-with-other-content.992578/

Repulsive sites in the midplane $z=0$ are arranged on the corners and edge centers of a square of side length $2 \sigma$, the repulsive interaction has range $2^{1/6} \sigma$, so the attractive sites are chosen to sit approximately $0.2 \sigma$ from the effective excluded volume of the monomer.
% Eq. (\ref{eq:repcoords}) defines particles with \GMG{face-to-face??} separation approximately $(0.4 + 2^{1/6}) \sigma$ in the target configuration, whereas the repulsive interaction potential goes to zero when repulsive sites have separation $2^{1/6} \sigma$. \GMG{(The following sentence is completely confusing to me as $i,j$ are integers.)} Alternatively, if $i,j$ were extended out to values of $d/2$, the repulsive volumes would overlap only at their boundary for $\phi_0 = 0$ and the surface of contact would reduce to a line of contact parallel to the common initial ${\bf c}_3 = {\bf c_1} \times {\bf c_2}$ for $\phi_0>0$ . %DMH  : this is my result from mathematica-aided algebra but I don't have a true lower bound for the excluder/repulsive site separation so no rigorous limit for how small deformations need to be. I could try to work more on this, or illustrate this "contact line" described in this sentence in fig a
Note that for $\phi_0 \neq 0$ the location of the repulsive sites do not lie along planar faces (as shown schematically in Fig.~\ref{fig:monomer_design}c), but more accurately instead skew quadralateral surfaces which deviate slightly from planarity. In Fig. \ref{fig:fig1a}, plots of pairwise energy under small deformations illustrate that repulsive sites do not play a role for the deformations up to the point of yielding (which can be roughly identified with the inflection point for energy vs. displacement), and for the parameters used in this work.

The interaction between attractive sites of the same type is given by the pair potential defined in Eq. \ref{eq:attractive} and there is no interaction between sites of different type.
The 8 types at different locations (see Fig. \ref{fig:monomer_design}) are used to control monomer face-face binding orientation.  Based on the form of Eq. \ref{eq:attractive}, the depth of each attractive potential pair is $-u_0/4$, so that an unfrustrated ($\theta_0=0$) bulk assembly has potential energy $-2 u_0$ per monomer (there is energy $-u_0$ per bond in the assembly, consisting of four attractive site pairs). There are four attractive sites on each side of the monomer (arrayed on a common plane), with corresponding types on opposite sides (see attractive sites colored by type in Fig. \ref{fig:monomer_design}b-c).

% Unlike the repulsive sites, t
The attractive sites on each side are arrayed in a strictly planar configuration, which is not generally the case for the repulsive sites arranged according to Eq. \ref{eq:repcoords}. For the side binding with respect to the ${\bf c_\alpha}$ direction, the four attractive sites have center of mass at $\frac{1}{2}d {\bf c_\alpha}$ with respect to the monomer center. From the center of mass, each attractive site is displaced a distance of $ \approx t = \sigma$ in either the $\pm {\bf c}_3'$ (i.e. vertically on the face) or $\pm {\bf c_\alpha}' \times {\bf c}_3'$ (i.e. horizontally on the face), where ${\bf c}_3'$ and ${\bf c_\alpha}'$ are the monomer's coordinate frame after rotation about ${\bf r}_\alpha$ by $\theta_0/2$.

The two attractive sites per side that are activated by both bending and twist deformations discussed below, displaced ``vertically'' from the center of each binding face,  are defined using the monomer coordinate frame ${\bf c}_1, {\bf c}_2, {\bf c}_3$ by 
\begin{widetext}
\begin{multline}
{\bf R}_{i,j}^{\rm vert} = \Big[ 
 \frac{d}{2} \cos \left( \frac{i \pi }{2} \right) + \sigma (-1)^i (-1)^j \tan \left(\frac{\theta_0}{2}\right) \cos \left(\frac{\pi  i}{2}+2 \phi_0 \right)\Big] {\bf c}_1  \\
 +\Big[\frac{d}{2} \sin \left( \frac{i \pi }{2} \right) + \sigma (-1)^i (-1)^j \tan \left(\frac{\theta_0}{2}\right) \sin \left(\frac{\pi  i}{2}+2 \phi_0\right)\Big] {\bf c}_2   +\sigma (-1)^j {\bf c}_3
 %\rangle
,
\label{eq:vertcoords}
\end{multline}
\end{widetext}
for $i = 1..4$ indexing the four bonding directions and $j = 1,2$ referencing sites above/below the monomer midplane. From the coordinates above, the magnitude of the displacement from the midplane is $t_{\rm vert}/2 = \sigma 2 \sqrt{1 + \tan^2( \theta_0/2)}$. 

The attractive sites activated with twisting rotation about the bond axis but not bending rotations about the orthogonal direction, displaced ``horizontally'' from the center of the binding face, are defined by 
\begin{eqnarray}
{\bf R}_{i,j}^{\rm horiz} &= \Big[
 \frac{d}{2} \cos \left( \frac{i \pi }{2} \right) - \sigma (-1)^i (-1)^j \sin \Big(\frac{\pi  i}{2} \Big)\Big] {\bf c}_1 \nonumber\\
&+ \Big[ \frac{d}{2} \sin  \left( \frac{i \pi }{2} \right) + \sigma (-1)^i (-1)^j \cos \left(\frac{\pi  i}{2} \right) \Big] {\bf c}_2 \nonumber\\
&+ \sigma (-1)^{1 + i} (-1)^j  \sin(2 \phi_0)\tan \Big( \frac{\theta_0}{2} \Big)
{\bf c}_3%\rangle
,
\label{eq:horizcoords}
\end{eqnarray}
for $i = 1..4$ indexing bond directions again and $j = 1, 2$ indexing sites displaced in opposite directions on the same monomer face. From these coordinates, the displacement of these attractors from the binding face center is $t_{\rm horiz}/2 = \sigma \sqrt{1 + \sin^2 (2 \phi_0) \tan^2 (\theta_0/2)}$. 

While our elastic theory treats each monomer as perfectly rigid, SWM is enforced via harmonic potentials of the form $\frac{1}{2} k_\mathrm{spring} (r - r_0)^2$ where $k_\mathrm{spring}$ is the spring constant and $r_0$ is the rest length to maintain site separations according to the geometry above. Springs are applied between nearby repulsive sites, for all pairs with indices $i_1,i_2,i_3$ such that each index differ by at most one (i.e. springs between nearest, next-nearest, and next-next nearest neighbors). Each attractive site has springs enforcing its distance to the six nearest repulsive sites. All springs are kept at the same spring constant, which is incremented during energy minimization as detailed in Sec. \ref{sec:model_methods}.

The attractive site positioning described above determines the relative cost of different elastic deformations of the assembly. In Fig. A\ref{fig:fig1a}, the potential energy of a bound pair of monomers is plotted for deformations away from the optimal (target) pair configuration with energy $- u_0$, corresponding to all four attractive sites in contact with respective sites on the other monomer. Compression or stretching deformations, due to center-of-mass displacement by a distance $\Delta r$ that equally affects all four attractive sites, change the potential energy to leading order by $\frac{1}{2} k_\mathrm{stretch} \Delta r^2$ with $k_\mathrm{stretch} = 4 \partial_r^2 u(r) = {\pi^2 u_0}/{2 r_a^2} $. Bending rotation, of an angle $\Delta \theta$ away from the preferred bend angle $\theta_{\parallel}$ and about an axis through the horizontal attractive sites, displaces vertical attractive sites by a distance $\Delta \theta t_{\rm vert} / 2$ and changes the potential energy of the configuration to leading order by $\frac{1}{2} k_{\parallel} \Delta \theta^2$ with $k_{\parallel} =  {\pi^2 t_{\rm vert}^2 u_0}/{16 r_a^2}$. Twisting rotation of an angle $\Delta \theta$ away from the preferred twist changes the potential energy by $\frac{1}{2} k_{\perp} \Delta \theta^2$ with $k_{\perp} = {\pi^2 (t_{\rm  vert}^2+t_{\rm horiz}^2) u_0}/{16 r_a^2}$ due to displacement of vertical and horizontal attractive sites. These three pairwise elastic constants determine the continuum model elastic constants relevant to stress accumulation in the assembly, discussed in the next section on the continuum theory. These constants depend weakly with $\theta_0$ and $\phi_0$ via $t_{\rm vert}$ and $t_{\rm horiz}$, which range from $t_{\rm vert}, t_{\rm horiz} = 2\sigma$ up to $t_{\rm vert}, t_{\rm horiz} \approx 2.008 \sigma$ for $\theta_0 = 10^\circ$, so that it approximately holds that $t_{\rm vert} \approx t_{\rm horiz} \approx 2 \sigma$. We therefore use a single elastic thickness parameter $t \equiv 2 \sigma$ independent of monomer shape parameters.

\section{Continuum elastic model for anisotropic membranes} \label{section:si_continuum}

In this section, continuum theory predictions of wedge assembly energetics are derived, namely the harmonic elastic flattening energy $E_\infty(\phi_0)$, the harmonic elastic energy of ribbons and rings $E(w, \phi_0)$, characteristic shape-flattening size $w_*(\phi_0)$ and numerical solutions for zero-temperature escape size $w_\mathrm{max}(\phi_0)$. These show the role of anisotropy of the assembly, as presented along with the discrete-SWM numerical results in the main text: the in-plane stretching cost depends only on a single elastic modulus $Y$ whereas the out-of-plane bending and the range of size control are dependent on the anisotropic bending costs $B_\parallel, B_\perp$ and preferred direction of curvature $\phi_0$. 

The continuum model presented here follows previous theory that captured flattening of frustrated membranes \cite{Ghafouri2005, armon2011geometry, grossman2016elasticity} and the equivalent description of shape selection in frustrated elastic sheets. \cite{jeon2017reconfigurable} The model predictions relevant to this study, including stress accumulation in the self-limiting regime (narrow limit), flattening transition and flattening cost are captured in an approximate description of the assembly geometry described by a single curvature tensor with components $C_{xx}, C_{yy}, C_{xy}$. This is an approximation to the shape of either slender ribbon or ring assembly, both with translational symmetry along the midline, and it is exact for the tubule-shaped membrane which is achieved in the limit of flattening $w \to \infty$ without yielding. A correction to the theory that accounts for varying curvature throughout the assembly, in the closed-ring case $\phi_0=0$, is detailed at the end of this appendix. To map the discrete wedge model to the continuum model, we additionally assume wedge monomer orientations stay aligned with the surface, so the monomer frame direction ${\bf c}_3$ is aligned everywhere with the local surface normal of the assembly. 

The elastic energy of the assembly is partitioned into two terms, applicable for the wedge assembly in the case of small deformations when the harmonic approximation to the attractive potential results in separate terms for stretching, bending and twist deformations of wedge bonds derived in the previous section. The continuum elastic energy is 
\begin{eqnarray}
\label{eq: elastic}
E_\mathrm{elastic} = E_{\rm strain} + E_\mathrm{bend} 
\end{eqnarray}
describing respective elastic costs due to monomer spacing deviating from $d$ and gradients in monomer orientations  deviating from the preferred angles $\theta_{\parallel}$ and $\theta_{\perp}$. 

The stretching cost is derived from the in-plane response to given curvatures $C_{xx}, C_{yy}, C_{xy}$ and the monomer preferred configuration, i.e. a square lattice with preferred spacing $\approx d$. For a 2D material with square symmetry, the full elastic energy has the form
\begin{multline}
E_\mathrm{strain} = \int dA \, \Big\{ \frac{\lambda}{2} (u_{xx} + u_{yy})^2  + \mu (u_{xx}^2 + 2 u_{xy}^2 + u_{yy}^2) \\ + \lambda_{\perp} u_{xx} u_{yy} \Big\},
\label{eq:square_stretch}
\end{multline}
where $u_{ij}$ is the strain tensor and $\lambda, \mu, \lambda_{\perp}$ elastic constants. \cite{LIFSHITZ19861} The corresponding stress tensor $\sigma_{ij}$ satisfying that $E_\mathrm{strain} = \int dA \sigma_{ij} u_{ij}$, is related to the strain by 
$u_{xx} = ({1}/{Y})\Big(\sigma_{xx} - \sigma_{yy}({\lambda + \lambda_{\perp}})/({\lambda + 2 \mu}) \Big), 
u_{yy} = ({1}/{Y})\Big(\sigma_{yy} - \sigma_{xx}(\lambda + \lambda_{\perp})/(\lambda + 2 \mu) \Big),$ and $\sigma_{xy} =  u_{xy} / (2 \mu)$ where 
\begin{eqnarray}
Y \equiv \frac{(2\mu - \lambda_\perp)(2 \lambda + 2 \mu + \lambda_\perp)}{\lambda + 2 \mu}
\end{eqnarray}
is the Young's modulus measured upon loading parallel to either of the square lattice close-packed directions. 
The strain depends on out-of-plane deflections, via $u_{ij} = \frac{1}{2}(\partial_i u_j + \partial_j u_i + \partial_i f \partial_j f )$ where $u_i$ is the in-plane displacement field and $f(x,y)$ is the deflection out-of-plane. \cite{Seung1988} A condition for a single-valued displacement field, the compatibility condition, is derived from applying $\epsilon_{ik} \epsilon_{jl} \partial_k \partial_l u_{ij}$ to the relations of strain to both the stress and the displacements. The special case of translational symmetry in the $y$ direction, so that gradients in $y$ vanish, implies $\sigma_{xx}, \sigma_{xy}$ are constants by mechanical equilibrium $\partial_i \sigma_{ij} = 0$, and compatibility becomes
\begin{eqnarray}
\epsilon_{ik} \epsilon_{jl} \partial_k \partial_l u_{ij} =  \frac{1}{Y}\partial_x^2\sigma_{yy} = -K_G
\end{eqnarray}
where $K_G = C_{xx} C_{yy} - C_{xy}^2 \simeq \det \partial_i \partial_j f$ is the Gaussian curvature, with the relation to $f(x,y)$ under the small-slope approximation. One finds the constants of integration that fully determine $\sigma_{ij}$ by minimizing over $E_\mathrm{strain}$. Thus, the continuum model expression for stretching cost of slender ribbons or rings of length $L$ and narrow width $w$ is 
\begin{eqnarray} \label{eq:si_estrain}
\frac{E_\mathrm{strain}}{w L} = \frac{Y}{1440} ( C_{xx} C_{yy} - C_{xy}^2)^2 L w^5.
\end{eqnarray}
We note that different prefactors appear in the same ``narrow-ribbon" constant curvature calculations in other Refs. \cite{grossman2016elasticity, meng2014elastic, Zhang2019, majidi08}, with specific Poisson ratio dependence, while our result in eq. (\ref{eq:si_estrain}) is inagreement with the (Poisson ratio-indendent) results of Refs. \cite{Ghafouri2005, Schneider2005}.  We believe this discrepancy to derive from the neglect of vanishing longitudinal net stress along the ribbon (i.e. $\int_{-W/2}^{+W/2} dx~ \sigma_{yy}(x) =0$) in the former group of references.  The fit-free agreement of numerical results presented in Fig. ~\ref{fig:flattening} with the eq. (\ref{eq:si_estrain}) is consistent with this conclusion.

The Young's modulus $Y$ is the response to uniaxial stress applied in $x$ or $y$ coordinate directions, which are the bonding directions for the wedge model. The stretching response of the assembly then derives from the pairwise stretching of bonds described in the previous section, so that taking the continuum limit of a square lattice of springs with spacing $d$, and extension along the bond direction, we have form of the the Young's modulus given in eq. (\ref{eq: params}), $Y = k_\mathrm{stretch}$.

The harmonic bending energy is also anisotropic. Tme most general form for the harmonic bending energy of membranes with anisotropy arising from distinct in-plane directions is given 
\begin{multline}
E_\mathrm{bend} = \int dA \Big\{\frac{1}{2} B_{xx} (C_{xx} - C_{xx}^0)^2 + \frac{1}{2} B_{yy} (C_{yy} - C_{yy}^0)^2  \\ +B_{xy}(C_{xy}-C_{xy}^0)^2   + \frac{1}{2}{\bar B}(C_{xx}C_{yy}-C_{xy}^2)\Big\},
\end{multline}
 equivalent to the most general form given in Helfrich and Prost \cite{helfrich1988intrinsic}, where it was emphasized that $C_{xy}^0 \neq 0$ arises from molecular chiral asymmetry despite the achiral symmetry of the curvature quadratic form. The case of $C_{xx}^0 = C_{yy}^0$, with zero Gaussian curvature modulus ${\bar K}=0$ was considered in Ref. \citenum{Ghafouri2005} to explain chiral amphiphile assemblies. The more general case with varying $C_{xx}^0, C_{yy}^0$ was connected to internally stressed elastic solids in Ref. \citenum{armon2011geometry}, where $C_{xy}^0 \neq 0$ could be realized from stretched bilayers without material chirality, instead the assignment of $x,y$ via the boundary breaks symmetry and results in chiral shapes. The interaction specificity in the wedge model distinguishes bonding in the two lattice row directions, so wedge monomers are notably chiral for the case $\phi_0>0$. If the attractive sites were arranged differently on the $x-$ and $y-$ oriented faces, the monomers could in principle have $B_{xx} \neq B_{yy}$. The saddle-wedge design presented in this study has $B_{xx}=B_{yy}=B_{\parallel} \neq B_{xy} = B_{\perp}$, with consequences for size control derived below. The role of Poisson's ratio in the elastic sheet description of Refs. \citenum{grossman2016elasticity, armon2011geometry} can be mapped to the Helfrich-Prost bending energy functional of the assembly above, effectively modulating both $B_\parallel/B_\perp$ and $C_{xx}^0, C_{yy}^0$. Here, we further develop the analysis with varying $\phi_0$ but constant preferred principal curvatures set by $\kappa_0$.

The bending elastic cost derives from the monomer attractive site interactions and the target pairwise configuration derived in the previous appendix. Assuming monomers have their frame direction ${\bf c}_3$ aligned with the assembly surface normal, the deviation from preferred bend angle in the $x$ direction is taken to be $d \times C_{xx} - \theta_0^{\parallel}$, in the $y$ direction $d \times  C_{yy} + \theta_0^{\parallel}$ and the deviation of twist angle in either direction is $d \times  C_{xy} - \theta_0^{\perp}$.
The cost of bend deformations for a single SWM with these local curvature values, is then the sum of potential energy from displacements of the two vertical attractive sites in either direction and of the four horizontal attractive sites
\begin{multline}
\label{eq: bendSWM}
E^{\rm (wedge)}_\mathrm{bend}  =   2 u\Big(\frac{1}{2}t_{\rm vert} \sqrt{(d ~ C_{xx} - \theta_0^{\parallel})^2 + (d ~ C_{xy} - \theta_0^{\perp})^2}\Big)   \\
 + 2 u\Big(\frac{1}{2}t_{\rm vert} \sqrt{(d ~ C_{yy} + \theta_0^{\parallel})^2 + (d ~ C_{xy} - \theta_0^{\perp})^2}\Big)  \\+ 4u\Big(\frac{1}{2} t_{\rm horiz} (d ~ C_{xy} - \theta_0^{\perp})\Big) + 2 u_0 .
\end{multline}
where the attractive potential is, again, $u(r) = -\frac{1}{8}u_0\Big[1 + \cos \Big( {\pi r}/{r_a}\Big) \Big]$ when $r < r_a$. Summing over all wedges (per unit area $d^2$) and assuming small deformations, $u(r) \approx -\frac{1}{4} u_0 + \pi^2 u_0 r^2/(16 r_a^2)$ and the bending energy takes the (harmonic) form of Eq. (\ref{eq: bend}) with the moduli given in Eq. (\ref{eq: params}).  Thus, the SWM design has anisotropic bending constants, with $B_{\perp} = 2 B_{\parallel}$ arising from the configuration with two attractive sites displaced under bending and all four attractive sites displaced under twisting deformation.

%\begin{eqnarray} 
%E_\mathrm{bend} \approx \int dA \, &\frac{1}{2}B_{\parallel} \Big( (C_{xx} - \kappa_0 \cos(2 \phi_0))^2 + (C_{yy} + \kappa_0 \cos(2 \phi_0))^2)  \nonumber \\
%&+ B_{\perp} (C_{xy} - \kappa_0 \sin(2 \phi_0))^2 \Big),
%\label{eq:ebend_harmonic}
%\end{eqnarray}
%where 
%\begin{eqnarray}
%\kappa_0 = \theta_0/d, \, B_{\parallel} = B_{\parallel} \approx %\frac{\pi^2 t^2 u_0}{4 r_a^2}, \, B_{\perp} = B_{\perp} \approx %\frac{\pi^2 t^2 u_0}{2 r_a^2}.
%\end{eqnarray}

\begin{figure*}[h!]
% \centering\includegraphics[width=1.0\linewidth]{fig_sla_curves.jpg}
\centering\includegraphics[width=0.6\linewidth]{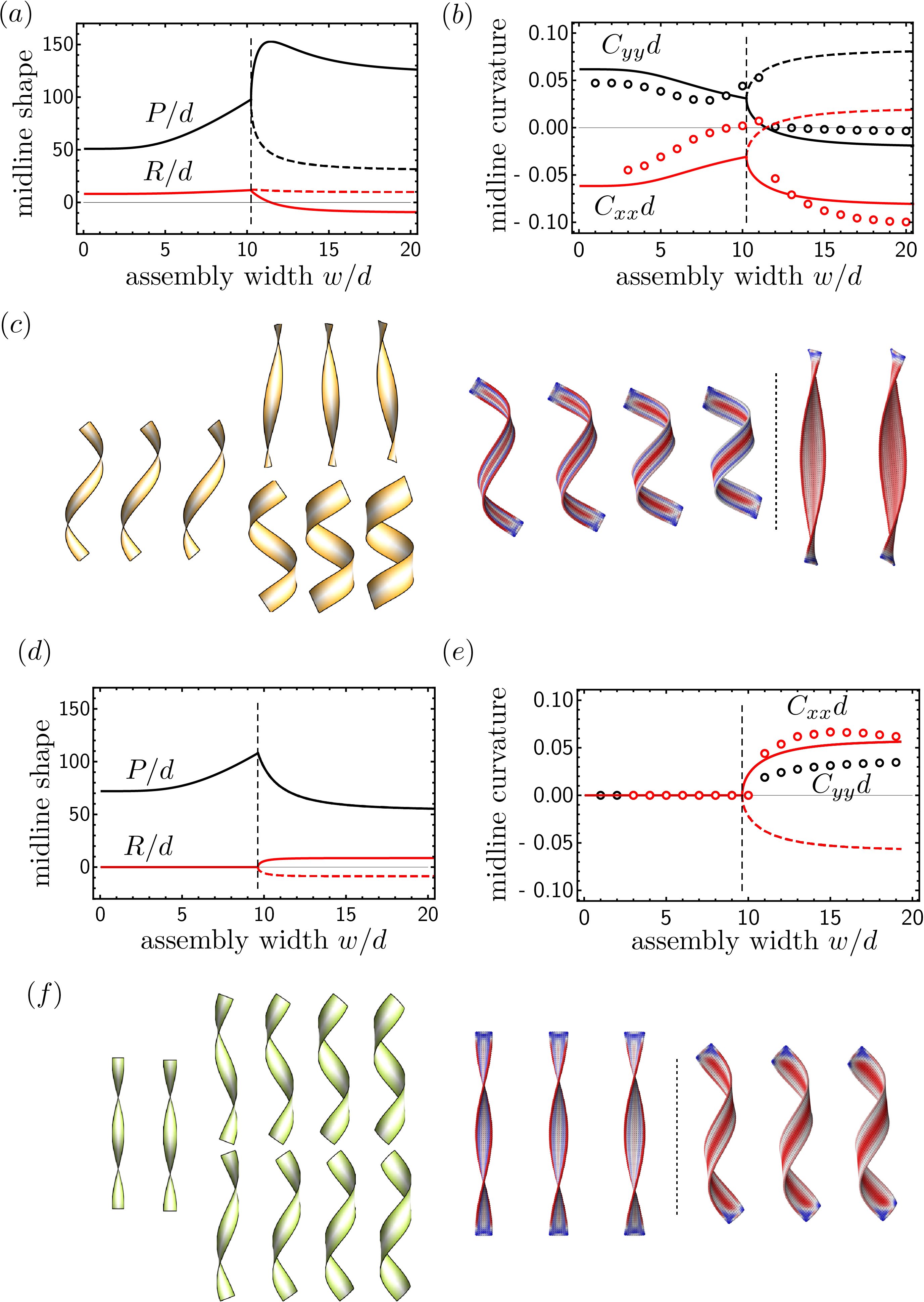}
\caption{  Two flattened shape minima are predicted in the wide branch of ribbon minima. Theory prediction and numerical results are shown for the case $\theta_0=5^\circ$ for centerline helical pitch $P$ and radius $R$ in (a) and (d), and centerline curvatures $C_{xx}, C_{yy}$ in (b) and (e) both pre- and post-flattening extracted from simulated SWM ground states. Dashed curves denote secondary branches that are of equal energy in the approximate theory. 
For the case $\phi_0=22.5^\circ$ shown in (a-c), corrections to the theory predict in Ref. \citenum{efrati2009buckling} predict that the dashed, smaller-pitch branch will be of lower energy. Numerical results (points) consistenly land in the higher-pitch shape solution. 
For the case $\phi_0 = 45^\circ$, shown in (d-f), the two flattened shapes are of equal energy, pitch and handedness but related by rotation about the helical axis $R \to -R$.  } \label{fig:fig_flattened_branches}
\end{figure*}

The NR theory elastic cost of the flattened state $E^{\infty}$ is found by optimizing $E$ under the constraint that the curvatures describe a cylindrical surface with zero Gaussian curvature, $C_{xx} C_{yy} - C_{xy}^2 = 0$. Thus, monomers arrange with preferred spacing $d$ and there is zero stretching energy so $E = E_\mathrm{bend}$ (minimized over subject $K_G =0$). Minimization $E_\mathrm{bend}$ subject to $K_G =0$ leads to two degenerate solutions, shown Fig. \ref{fig:fig_flattened_branches} for $\phi_0=22.5^\circ$ and $45^\circ$. In the harmonic approximation described by equation \ref{eq: bend}, one finds that both the optimal shape and energy depend on both $\phi_0$ and the twist-bend anisotropy. When $B_{\parallel}=B_{\perp}$ (as studied in the case of Ref. ~\cite{Armon2014}), one of two equal-energy shape solutions is $C_{xy}/\kappa_0 = \frac{1}{2}\sin(2\phi_0)$, $C_{xx}/\kappa_0 = \frac{1}{2}(\cos(2\phi_0)-1)$,  and $C_{yy}/\kappa_0 = -\frac{1}{2}(\cos(2\phi_0)+1)$, giving energy $E_\infty /A= \frac{1}{2} B_{\parallel} \kappa_0^2$ that is independent of $\phi_0$. 

For the present case of SWM assembly, were $2 B_{\parallel} = B_{\perp}$, one of the two shape solutions is 
\begin{eqnarray}
C_{xy}/\kappa_0 &=& \frac{2}{3}\sin(2\phi_0) \nonumber \\
C_{xx}/\kappa_0 &=& \frac{1}{2}\cos(2\phi_0) - \frac{1}{2}{\sqrt{\cos^2(2\phi_0) + 4 C_{xy}^2}}\nonumber \\ 
C_{yy} &=& -\frac{1}{2}\cos(2\phi_0) - \frac{1}{2}{\sqrt{\cos^2(2\phi_0) + 4 C_{xy}^2}}
\end{eqnarray}
corresponding to (Hookean elastic) continuum result for flattening energy,
\begin{eqnarray}
\frac{E_\infty}{A} = \frac{1}{2}B_{\parallel}\kappa_0^2 (1 + \frac{1}{3} \sin^2(2\phi_0)) , (\textrm{when } B_\perp=2 B_\parallel),
\end{eqnarray}
and more generally,
\begin{eqnarray}
\frac{E_\infty}{A} = \frac{1}{2}B_{\parallel}\kappa_0^2 \Big[1 + \frac{B_\perp - B_\parallel}{B_\perp + B_\parallel} \sin^2(2\phi_0)\Big].
\end{eqnarray}
This is the expression plotted in Fig.  \ref{fig:flattening}b for the harmonic limit $\theta_0/r_a \rightarrow 0$. To get the corrected result accounting for non-linear strain softening of the potential, also plotted in Fig. \ref{fig:flattening}b, eq. (\ref{eq: bendSWM})  was numerically optimized over  $C_{xy}, C_{xx}$ with $C_{yy}=C_{xy}^2/C_{xx}$ to produce the curves with nonlinear bending energy and nonzero $\theta_0/r_a$.
% could mention that energy and shape both vary for nonlinear case as function of \theta_0/r_a

The full NR elastic energy for the SWM model is the sum of the narrow-ribbon limit $E_\mathrm{strain}$ and the harmonic approximation of $E_\mathrm{bend}$ evaluated with constant curvatures everywhere:
\begin{multline} 
\frac{E}{w L} =  \frac{Y}{1440} ( C_{xx} C_{yy} - C_{xy}^2)^2 w^4  \\
+ \frac{1}{2}B_{\parallel} \Big[ (C_{xx} - \kappa_0 \cos(2 \phi_0))^2 + (C_{yy} + \kappa_0 \cos(2 \phi_0))^2\Big]  \\
+ B_{\perp} (C_{xy} - \kappa_0 \sin(2 \phi_0))^2,
\end{multline}
which is then optimized over the curvatures $C_{ij}$ for varying $w$. 

When $B_{\parallel} = B_{\perp}$, the branches of the equilbriumsolution are characterized by a length $w_*  = \Big[ 2880 {B_{\parallel}}/({Y \kappa_0^2}) \Big]^{1/4}$ associated with the transition from unflattened shape at small $w$ with zero mean curvature to one of two flattened branches of degenerate energy with analytical forms given in Ref. \citenum{grossman2016elasticity} up to a constant arising from different approximations for stretching energy. One can verify that the energy in both narrow and wide branches is independent of $\phi_0$. 
In general, we find
\begin{eqnarray} 
w_* = \Big[ \frac{ 2880 B_{\parallel} ( B_\perp + B_\parallel)^2}{Y \kappa_0^2 (( B_\perp + B_\parallel)^2 + (B_\perp - B_\parallel)(3 B_\perp + B_\parallel)\sin^2 (2\phi_0)) }\Big]^{1/4},
\end{eqnarray}
and the specific presnet case of the SWM model with anisotropic curvature moduli,
\begin{eqnarray} 
\label{eq: wstar}
w_* = \Big[ \frac{ 2880 B_{\parallel}}{Y \kappa_0^2 (1+\frac{7}{9}\sin^2 (2\phi_0)) }\Big]^{1/4}, \ \ \ \  \textrm{for }B_\perp=2 B_\parallel
\end{eqnarray}
so that increasing $\phi_0$, i.e. increasing preferred twist curvature, results in the flattening at a smaller characteristic width. The energy of the wide branch has the form,
\begin{eqnarray} 
%E_\mathrm{wide} = E_\infty \Big(1 - \frac{25 - \cos(4 \phi_0)}{12(6+2\sin^2 (2 \phi_0)) (w/w_*)^4} \Big), w > w_*.
\frac{E_{\rm elastic}(w\geq w_*)}{A} = \frac{E_\infty}{A} - \frac{360 B_\parallel^2}{Y w^4} , \ \ \ \  \textrm{for }B_\perp=2 B_\parallel .
\end{eqnarray}

The wide branch approaches the harmonic approximation to the flattening energy $E_\infty$ that itself depends on $\phi_0$. The decay of the residual energy $\propto w^{-4}$ has the consequence that there is no surface energy $\gamma$ such that the model energy density $U/A = E(w)/A + 2\gamma/w$ has a minimum in the wide branch. Because of this, $w_*$ is strict upper bound for the optimal size $w_0$ that can be obtained from this theory, and the analytical expression for $w_*$ decreases with $\phi_0$ despite the increasing flattening cost with $\phi_0$ for this model. 

\begin{figure*}[]
\centering\includegraphics[width=1.0\linewidth]{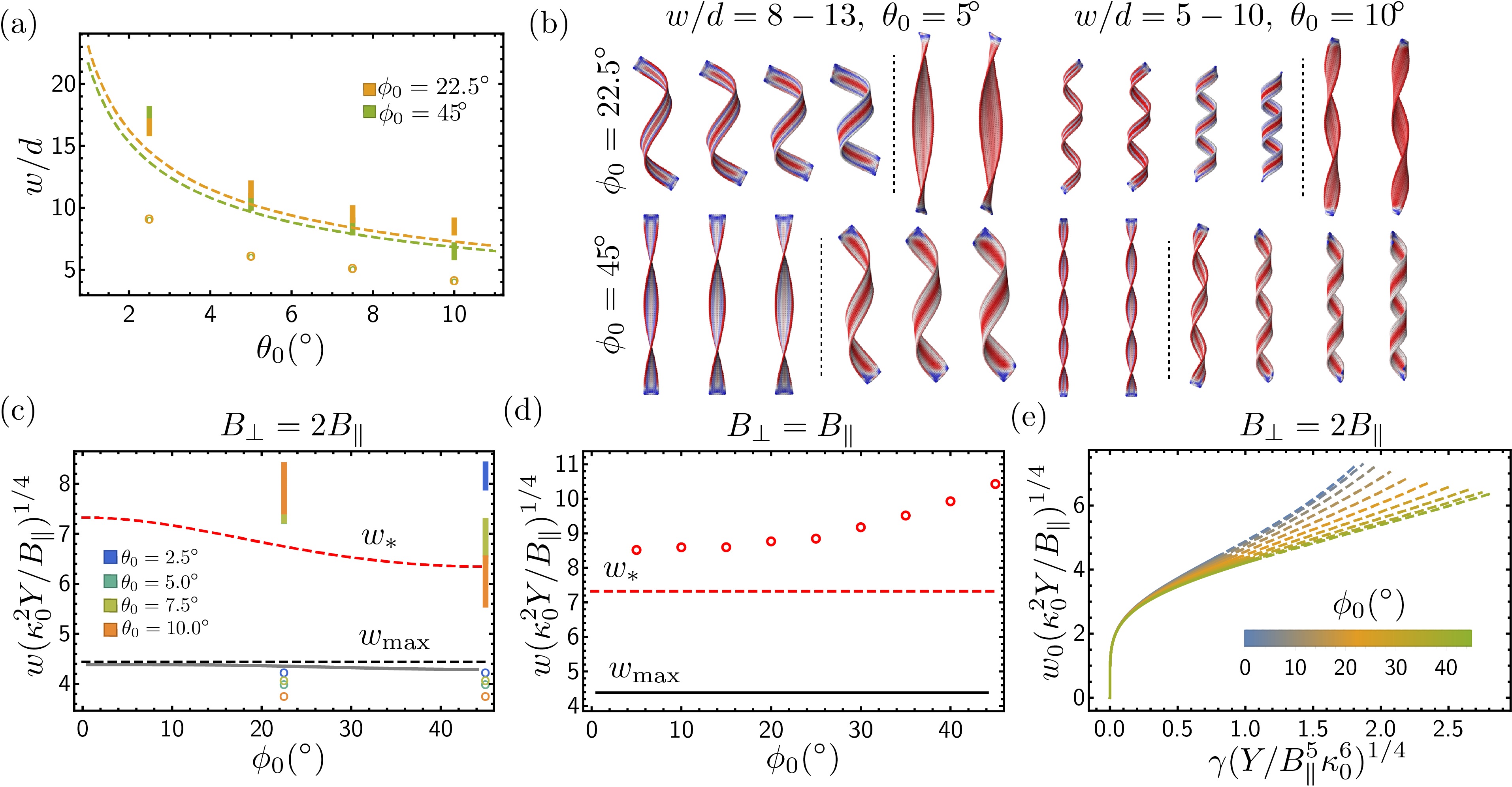}
% \centering\includegraphics[width=1.0\linewidth]{fig1.png}
\caption{ 
%Continuum theory prediction for the flattening transition size $w_*$ and  maximum size of equilibrium self-limitation $w_{\rm max}$.%, for the case $2 B_\parallel = B_\perp$. 
(a) Theory predictions (curves) are compared to numerical results (points) on the apparent critical width of the shape transition $w_*$. (b) Sequences of numerically minimized structures are shown, where numerical estimates of $w_*$ (vertical dashed line) are determined from the apparent change in structure based on these visualizations. 
(c) The solid black curve is the maximum size predicted by the continuum theory with shapes restricted to uniform curvature, rescaled by material parameters, $w_{\rm{max}}(B_\parallel / Y \kappa_0^2)^{1/4}$. The dashed line is the correction for $\phi_0=0$ from allowing curvature to vary along the ring width. The characteristic flattening size $w_*$ is an upper bound to the maximum size, plotted as a dashed red curve.  Open circles are $w_*$ results from finite element numerics taken from Ref. \citenum{Armon2014}. 
(d) The corresponding curves for the theory when $B_\parallel = B_\perp$ have no dependence on $\phi_0$ as a consequence of the isotropic flattening cost. Numerical results from Ref. \citenum{Armon2014} are plotted as open circles to show the magnitude of the affect that the authors attribute to the role of the boundary layer, supported by scaling arguments in Ref. \citenum{efrati2009buckling}. 
(e) Curves of constant $\phi$ are plotted for the equilibrium size $w_0$ as a function of line tension $\gamma$. The curves are plotted as solid up until $w_{\rm{max}}$, when the minimum at finite $w$ becomes unstable to the bulk tube energy $E_\infty$. The dashed curves track metastable $w_0$ up until $w_*$.}
\label{fig:contin_wmax}
\end{figure*}

Local minima exist for any $w<w_*$ in the narrow branch, with a discontinuity in the second derivative of $E(w)$ present at $w=w_*$. In the narrow branch, the maximum self-limiting size $w_\mathrm{max}$ is taken to be the minimum to $U/A = E(w)/A + 2\gamma/w$ at which the local minimum has equal energy to the bulk tube state.
%Importantly, the functional form of the wide branch has additional significance that, for any cohesive energy of the form ${F_\mathrm{coh}}/({w L}) = {\gamma}/{w}$ as discussed in the main text, there can be no value of $w$ that is stable to local growth and also of lower free energy than the bulk $w \rightarrow \infty$ state, 
That is, following the framework in Ref. \citenum{hagan2021}, self-limiting size is achievable when the {\it accumulant} is increasing, $\partial_w {\mathcal A} = \partial_w (w(E_\infty - E(w))/A) > 0$. This is found to always occur at a size smaller than $w_*$ in the model. The solution $w_\mathrm{max}$ satisfying $\partial_w {\mathcal A}=0$ from numerically solving for the narrow branch is plotted in Fig.  \ref{fig:contin_wmax}.  
% don't have analytical expressions for narrow branch E(w) nor $w_\mathrm{max}(\phi_0)$.

In general, a more accurate description of shape solutions with varying curvatures is expected to include a boundary layer that develops as $w \to \infty$, so the residual energy of wide ribbons $E_\mathrm{boundary} \sim 1/w$ allows for locally optimal sizes at any width. The calculation at the end of this section, however, shows that the correction in $w_\mathrm{max}$ is small for $\phi_0 = 0$. The boundary layer correction may have an effect on the dependence of $w_{\rm max}$ with $\phi_0$ : even for the isotropic case $B_\perp=B_\parallel$, the boundary layer length was shown to scale depending on the tangential curvature of the flattened ribbon $l_{\rm boundary} \sim \sqrt{ \sqrt{B/Y} / C_{yy}}$ in Ref. \citenum{efrati2009buckling}, and as a consequence $w_*$ increases with increasing $\phi_0$ supported by finite-element numerical results in Ref. \citenum{armon2011geometry}. The weak dependence of $w_\mathrm{max}$ on $\phi_0$ in the numerics presented in this study may be affected by the effect captured in this boundary layer scaling : increasing $\phi_0$ decreases tangential curvature $C_{yy}$ along the boundary of the flattened ribbon. Whereas the initial elastic energy growth $E \sim Y \kappa_0^4 w^4$ depends only on the intrinsic geometry, the extrinsic geometry, $\phi_0$, can affect the energetics, and range of equilibrium size control, via the mechanics at the boundary. 
% \DMH{ went through Eran Sharon references and the boundary layer length scaling seems to be the most relevant result, to influence phi-dependent effects on $w_*,w_\mathrm{max}$.}

% not sure what to cite and how much to include regarding trumpets
%also wondering if this works almost immediately without much modification for phi>0, seems like it should just expanding for a post-flattened, cylindrical Monge patch with appropriate curvatures?
For the case of the $\phi_0 = 0$ ring assembly we derive the boundary layer corrected solution starting from harmonic elastic energy described by equations \ref{eq:square_stretch} and \ref{eq: bend}, limited to axisymmetric shapes, i.e. translation along the $y$ coordinate. Here, the optimal shape is allowed to adopt in-plane displacement $u_x(x)$ transverse to the ring and out-of-plane displacement $h(x)$ in the radial direction, expanding around a cylindrical shape with principal curvature $C_{yy} = -\kappa_0$ with linearized solution for $h(x) \kappa_0 \ll 1$. One finds that  
\begin{eqnarray}
\label{eq: BLcat}
C_{yy} & \simeq& -\kappa_0 + \kappa_0^2 h \\
C_{xx} & \simeq& \partial_x^2 h \\
C_{xy} &=& 0\\
u_{xx} &= &\partial_x u_x + \frac{1}{2} \Big((\partial_x h)^2 + (\partial_x u_x)^2 \Big)\\
u_{xy} &=&0\\
u_{yy} &=&\kappa_0 h + \frac{1}{2}\kappa_0^2 h^2\\
\end{eqnarray}
and defining characteristic length $\lambda$:
\begin{eqnarray}
\label{eq: excesscat}
\lambda &\equiv& \Big(\frac{4 B_{\parallel}}{Y \kappa_0^2} \Big)^{1/4}\\
E_\infty &=& \frac{1}{2} B_{\parallel} \kappa_0^2 w L\\
E_{\rm elastic} &=& E_\infty \Big( 1 +2( \lambda/w)\frac{\cos(w/\lambda) - \cosh(w/\lambda)}{\sin(w/\lambda) + \sinh(w/\lambda)}\Big)\\
\frac{E}{w L} &\simeq& \frac{1}{1440}Y \kappa_0^4 w^4 , (\textrm{when }w \ll \lambda).
\end{eqnarray}
Where the $w\to 0$ limit on the last line is in agreement with the result of the uniform curvature approximation. Following the accumulant analysis of Ref. \citenum{hagan2021}, the maximum possible self-limited size becomes
\begin{eqnarray}
\label{eq: wmaxcat}
w_\mathrm{max}=\pi \lambda.
\end{eqnarray}
For $\phi_0 = 0$, $w_* =(720)^{1/4} \lambda \approx 5.18 \lambda$, and approximately $1.3 \%$ greater than $w_\mathrm{max}$ from the uniform curvature result, as shown by the comparison in Fig.  \ref{fig:contin_wmax}. In this way, corrections to the model results with the uniform curvature approximation are expected to be small for the energetics up to $w_\mathrm{max}$. A similar result for the case of $\phi_0=45^\circ$ was included in Ref. \citenum{arieli2021geometric}.

\end{document}